\documentclass[twocolumn,prb,superscriptaddress]{revtex4}
\usepackage{graphicx,float}
\usepackage{latexsym}
\usepackage{epstopdf}
\usepackage{amssymb,amsmath}
\usepackage{color}
\usepackage{multirow}
\usepackage{hyperref}

\makeatletter
\let\Hy@linktoc\Hy@linktoc@none
\makeatother

\usepackage{verbatim}
\usepackage{bm}

\DeclareMathOperator{\tr}{tr}

\begin{document}

\title{Quantum entanglement entropy and classical mutual information in long-range harmonic oscillators}
\author{M. Ghasemi Nezhadhaghighi}
\affiliation{Department of Physics, Sharif University of Technology, Tehran, P.O.Box: 11365-9161, Iran}
\affiliation{Institute for Physics \& Astronomy, University of Potsdam,14476 Potsdam-Golm, Germany, EU}
\author{ M. A. Rajabpour}\email{rajabpour@ursa.ifsc.usp.br}
\address{Instituto de F\'{\i}sica de S\~{a}o Carlos, Universidade de S\~{a}o Paulo, Caixa Postal 369, 13560-970 S\~{a}o Carlos, SP, Brazil}

\begin{abstract}
We study different aspects of quantum von Neumann and R\'enyi entanglement entropy of one dimensional 
long-range harmonic
oscillators that can be described by well-defined non-local field theories. 
We show that the entanglement entropy of one interval with respect to the rest changes logarithmically with the number of 
oscillators inside 
the subsystem. This is true also in the presence of different boundary conditions. We show that the coefficients of the logarithms coming from different 
boundary conditions can be reduced to  just two different universal coefficients. We also study the effect of the mass and temperature on the entanglement entropy of
the system in different situations. The universality of our results is also confirmed by changing different parameters in the coupled harmonic oscillators.
We also show that more general interactions coming from general singular Toeplitz matrices can be decomposed to our long-range harmonic oscillators. Despite the long-range
nature of the couplings we show that the area law is valid in two dimensions and the universal logarithmic terms appear if we consider subregions with sharp corners. Finally we
study analytically different aspects of the mutual information such as its logarithmic dependence to the subsystem, effect of mass and influence of  the boundary. We also 
generalize our results in this case to general singular Toeplitz matrices and higher dimensions.

\end{abstract}

\maketitle
\tableofcontents

\section{Introduction}\label{sec1}

Quantum entanglement entropy as an interesting quantity in many body systems has been studied in many different 
locally interacting systems by using different techniques, see the reviews \cite{amico2008,Modi2012,Eisert2010,Calabrese2009,Casini2009, Peschel2009} and references therein. 
Among the most
important results (which are related to this work) one can list the classical result of Bombelli, et.al \cite{Bombelli1986} which  they compute
the entanglement entropy of free field theory by using the discrete version of the field theory which is simply coupled harmonic oscillators.
The result was rediscovered in \cite{Srednicki1993} and used to introduce the area law. In \cite{Callan1994} the result was generalized to the R\'enyi entropy
and the validity of the replica trick is checked. This method is also used to study free fermionic systems in a series of papers by Peschel and
collaborators in \cite{Peschel2003,Peschel2009}. The techniques used in these works are applicable in any dimension. In two dimensions for the short-range
interacting systems one can also 
derive exact formulas for the entanglement entropy using Euclidean methods \cite{Casini2009}. In the especial cases when one have integrable models
one can use the form factor techniques and calculate the entanglement entropy \cite{Castro-Alvaredo2009}. Finally at the quantum critical point when we have conformal field theory 
many explicit results are known, see \cite{Calabrese2009} and references therein. 

Although in short-range interacting systems numerous results has been discovered in last ten years there are just few results concerning long-range interacting
systems. The main difficulty is the lack of exact solution in most of this kind of systems.  The entanglement entropy in Lipkin-Meshkov-Glick 
(LMG) model which in that all spins interact among themselves is studied numerically and analytically in \cite{Lattore2005,Orus2008}. In \cite{Dur2005}, 
  the static and the
dynamical  properties of the entanglement entropy is studied in a long-range Ising type model without an external magnetic field. In the same 
direction the entanglement entropy  is also calculated numerically for the anti-ferromagnetic long-range Ising chain in \cite{Koffel2012}.
In an interesting work a logarithmically divergent geometric entropy is found 
 in free fermions with long-range unshielded Coulomb
interaction in \cite{Eisert2006}. Plenio, et. all \cite{Plenio2005,Cramer2006, Audenaert2005}, see also \cite{Botero2004,Cramer2006b} studied the general properties 
of the entanglement entropy for coupled  harmonic
oscillators and found an interesting bounds for the entanglement entropy.  Finally using the matrix product states it was argued in \cite{Cadarso2013} 
 that for those long-range systems
that  one can not approximate the ground state of the model with the ground state of another short
range model, we expect larger entanglement. One can find some other results concerning entanglement entropy in long-range systems in \cite{Others}.

Recently using the methods of \cite{Bombelli1986,Callan1994} we studied the entanglement entropy of block of long-range coupled  harmonic
oscillators \cite{Nezhadhaghighi2012}. We showed that the entanglement of the gapless system is logarithmically dependent to the system size
and we calculated the prefactor of the logarithm in different situations. The idea of studying this particular non-local system is manyfold: firstly
the hamiltonian (\ref{harmonicOsc}) that we are going to study is a simple discretization of fractional laplacian and so it has a very
clean continuum limit. This is useful because then we can claim that we are actually studying the entanglement entropy of
a non-local field theory. This field theory is a well-known field theory which also appears in the study of long-range Ising model \cite{Blanchard2013} so
in principle any analytical understanding of the entanglement entropy of  long-range Ising model will be based on the system
that we are studying. Having the above motivations in mind we extended our study in many different directions. 

The organization of the paper is as follows: in section two we present the model and give the definitions of the quantities that
we are going to study. In section three we study different aspects of von Neumann and R\'enyi entanglement entropy in long-range harmonic oscillators.
We first summarize the main formulas that we need to calculate the entanglement entropy. Most of the formulas are in the discrete level
but we also provide the eigenvalue problems in the continuum limit. Then we study the entanglement entropy numerically both at the purely discrete
level and also at the level of discretization of the eigenvalue problem. This part of the paper is the extension of the work done in \cite{Nezhadhaghighi2012}.
Then we study the finite size effects in different kind of situations such as, periodic boundary conditions and Dirichlet boundary conditions. Then we
compare the results with the massive coupled long-range oscillators. After that we study the effect of temperature on the entanglement entropy 
of our system. Our main result will be presented at the end of this section which concerns the universality of our results. In this section
we will show that the results presented in the previous sections are robust against many small changes in the form of the interaction. We will also 
show that one can calculate the entanglement entropy of larger set of coupled oscillators, to be specific oscillators coupled
with singular Toeplitz interactions, to   the cases that we studied in previous sections. we will conclude this section with some comments about 
the entanglement entropy in higher dimensions especially in the presence of polygonal regions.
Finally in section three we will study different aspects of the classical mutual information in long-range harmonic oscillators. We will presents 
two definitions and then using Fisher-Hartwig theorem we will show that in contrast to the
von Neumann entanglement entropy one can actually analytically calculate these quantities. In this section we
also address the finite size effects and also the massive case. The generalization to the singular Toeplitz matrices will be also 
discussed.

\section{Definitions and settings}\label{sec1}

We start by describing the coupled harmonic oscillators, the perhaps simplest 
lattice model available to the research where the hamiltonian is a quadratic form:
\begin{equation}\label{harmonicOsc}
\mathcal{H}=\frac{1}{2}\sum_{n=1}^{N}\pi_n^2+\frac{1}{2} \sum_{n,n^\prime=1}^{N}\phi_{n} K_{nn^\prime}\phi_{n^\prime}~.
\end{equation}
We would like to study coupled harmonic oscillators with long-range interaction. 
To define the $K$ matrix for the long-range harmonic oscillator problem one can use the fractional operator. In principle there are many ways to write 
a long-range $K$ matrix, however, we are interested in those that they have a very simple continuum counterpart. In principle in the continuum
the fractional laplacian is usually defined by its Fourier transform $|q|^{\alpha}$ or $(q^2)^{\frac{\alpha}{2}}$ which $q^2$ is just the Fourier transform
of a simple laplacian. Since the fourier transform of the discrete laplacian is $2-2\cos q$ one may use some powers of this to define the
discrete fractional laplacian. Then the elements 
of the matrix $K$, representing the discretized fractional Laplacian, are  
\begin{eqnarray} \label{HOLR} 
\begin{split}
K_{l,m} &=-\int_0^{2\pi} \frac{dq}{2\pi} e^{iq(l-m)} \lbrace \left[ 2(1-\cos(q))\right]^{\frac{\alpha}{2}}+{M}^{\alpha}\rbrace \\ 
&= \frac{\Gamma(-\frac{\alpha}{2}+n)\Gamma(\alpha +1)}{\pi 
\Gamma(1+\frac{\alpha}{2}+n)} \sin(\frac{\alpha}{2}\pi)+{M}^{\alpha}\delta_{l,m}~, 
\end{split}
\end{eqnarray}
where $n=|l-m|$, and  fractional order $\alpha>0$. In the future $M$ will play the role of the mass of the fractional field theory. 
In the special case $\alpha=2$ 
the $K$ matrix is equal to the simple laplacian. When $\alpha/2$ is an integer 
the elements 
$K(n) = (-1)^{\alpha-n+1}C_{\alpha,\frac{\alpha}{2}+n}$ for $n\leq\alpha/2$ and 
$K(n)=0$ for $n>\alpha/2$, where $C_{\alpha,\frac{\alpha}{2}+n}$ are binomial coefficients \cite{Zoia2007}. 

For sufficiently large one dimensional system, $K$ and the two point correlator 
matrices $K^{\pm 1/2}$ are Toeplitz matrices and all off-diagonal elements of them are 
identical. The elements of $K^{\pm 1/2}_{l,m}$ can be expressed as a Fourier series

\begin{eqnarray} \label{HOLR2} 
\begin{split}
K^{\pm 1/2}_{l,m} &= K^{\pm 1/2}(n) = -\int_0^{2\pi} \frac{dq}{2\pi} e^{iq(l-m)}\times \\
&\lbrace \left[ 2(1-\cos(q))\right]^{\frac{\alpha}{2}}+{M}^{\alpha}\rbrace ^{\pm 1/2}~.
\end{split}
\end{eqnarray}
The matrix $K^{- 1/2}$ corresponds to the spatial correlation 
of an oscillator system $\langle \phi_l\phi_m \rangle$, and for the system with periodic
 boundary condition one can find the spatial correlation length $\xi_s$ as \cite{Unanyan2005}
 \begin{eqnarray}
 \begin{split}
 \xi_s ^{-1}&\equiv -\lim_{n\rightarrow\infty}\frac{1}{n}\log |\langle \phi_l\phi_{l+n} \rangle| \\
 &= -\lim_{n\rightarrow\infty}\frac{1}{n}\log |K^{- 1/2}(n)|.   
\end{split} 
 \end{eqnarray}
For the massless system $\xi_s ^{-1} = -\lim_{n\rightarrow\infty}\frac{1}{n}\log |\frac{\Gamma(n+\alpha/4)\Gamma(1-\alpha/2 )}{\pi \Gamma(1-\alpha/4+n)} \sin(\frac{\alpha}{4}\pi)|=0$ and for the massive case $\xi_s ^{-1} \propto M$. We note that for $M=0$, the correlation length $\xi_s$ is infinite and the system is gapless, and for non-zero value of $M$ the system is gapped.

The $K$ matrix in the continuum limit has the following form:
\begin{eqnarray} \label{fractional free field theory}
\begin{split}
&\frac{1}{2} \sum_{n,n^\prime=1}^{N}\phi_{n} K_{nn^\prime}\phi_{n^\prime}\rightarrow \\
 &\int \lbrace-\frac{1}{2}\phi(x)(-\nabla)^{\alpha/2}\phi(x)+\frac{1}{2}{M}^{\alpha}\phi^2(x)\rbrace dx,
\end{split}
\end{eqnarray}
where $-(-\nabla)^{\alpha/2}$ is defined by its Fourier transform $|q|^{\alpha}$.

We are now in a position to introduce the entanglement entropy and it's value in two dimensional CFT's. Here, we shall only
discuss the von Neumann and R\'enyi entanglement entropies. Nevertheless, there are many other measures that
have been explored \cite{amico2008,Eisert2010,Modi2012}. 

Consider a system with the density matrix of a pure state $\rho$, which is divided into two subsystems
 $A$ and $B$. Then the entanglement may be characterized by the properties of the reduced density matrix $\rho_A$
 of the subsystem $A$. Density matrix $\rho_A$ is obtained by tracing out the remaining degrees of freedom $\rho_A = \tr_B \rho$. 
The von Neumann entanglement entropy associated to the local
density matrix $\rho_A$ reduced to a region $A$ of the space is 
\begin{equation} \label{} 
S(A) = -\tr(\rho_A \log(\rho_A))~.
\end{equation}
Another related measure to the local density matrix, is a family of functions called the R\'enyi entropies,
\begin{equation} \label{Renyi} 
S_{n}(A) = \frac{1}{1-n}\log(\tr \rho_A^n), \hspace{0.5cm} n\geq 0, \hspace{0.5cm} n \neq 1~.
\end{equation}
 The R\'enyi entropy $S_n$ has similar properties as the entanglement entropy 
$S$. 

For general quantum field theories in $d$ spatial dimensions the entanglement entropy 
is always divergent in a continuum system and the coefficient of the leading divergence term
is proportional to the area of the boundary of the subsystem $A$ and it is given by the simple formula \cite{Casini2009}

\begin{eqnarray} \label{AreaLaw} 
\begin{split}
S(A) &= g_{d-1}\left(\frac{l}{\epsilon}\right)^{d-1}+ \dots+g_{1}\left(\frac{l}{\epsilon}\right)^{1}\\ &+g_{0}\log(l/\epsilon)+S_0(A)~,
\end{split}
\end{eqnarray}  
where $\lbrace g_{d-1},\dots,g_1\rbrace$ and $S_0$ are non-universal constants which depend on 
the system. The coefficient $g_0$ of the log term is expected to be universal
 and $l^d$ is the  volume in $d$ dimensional space and $\epsilon$ is a short distance cutoff (or a lattice spacing). 
The simple area law, however, can not describe the scaling of the entanglement entropy in generic 
cases. Indeed the entanglement entropy of conformal field theory in one special dimension, scales logarithmically with 
respect to the size of the subsystem  $l$. If the total system is infinitely long, it is given by the simple formula 
\begin{equation}\label{scalingEsmall_l}
S = \frac{c}{3}\log \frac{l}{\epsilon}~,
\end{equation}   
where c is the central charge of the CFT \cite{Holzhey1994}. In $1+1$ dimensional conformal invariant systems the Re\'nyi 
entropy follows  \cite{Calabrese2004}

\begin{equation}\label{scalingEsmall_l}
S_n = \frac{c}{6}(1+\frac{1}{n})\log \frac{l}{\epsilon}~.
\end{equation} 

It is also worth mentioning that for a finite system of length $L$ with boundary, 
at zero temperature and one special dimension, divided into two pieces of lengths $l$ and $L-l$, 
the R\'enyi entropy obeys
\begin{equation}\label{EE finite half}
S_n = \frac{c}{12}(1+\frac{1}{n})\log ((L/\pi a)\sin(\pi l/L))+c^\prime_1~.
\end{equation}  
 The above formulas are a few among many others that are known for different cases in two dimensional CFT's, see \cite{Calabrese2009}. 
In the next sections we will
introduce many of them as the limiting behavior of our long-range harmonic oscillators.

In the next section we will review a method where one can use it to calculate $\rho_A$ and consequently $S$ and $S_n$ for generic quadratic 
bosonic systems. Then we will hire this technique to study our particular long-range system. 

\section{von Neumann and R\'enyi entanglement entropy}
\subsection{Hamiltonian approach}\label{sec2}

A useful method to obtain entanglement entropy is introduced in \cite{Bombelli1986} and rediscovered in \cite{Srednicki1993}
and generalized to R\'enyi entropy in \cite{Callan1994}.  
In this method one would like to  measure the quantum entanglement entropy of the ground
state of the free field $\lbrace\phi \rbrace$, generated by tracing over fields inside of the region of the 
boundary surface. To fix the notation and for the later use we give here a brief summary of the work described in 
more detail in the Ref. \cite{Bombelli1986,Callan1994}. The ground state wave functional is given by 
 
\begin{equation} \label{GroundSwave} 
\Psi_0(\lbrace \phi \rbrace) \propto (\text{det}\Gamma)^{\frac{1}{4}}\exp \lbrace- \sum_{n,n^\prime=1}^{N}\phi_{n} 
\Gamma_{nn^\prime}\phi_{n^\prime} \rbrace.
\end{equation}
where $\lbrace \phi \rbrace$ denotes the collection of all $\phi$'s, one for each oscillator and $\Gamma = K^{1/2}$.
 
Now consider a subregion in the total space and split the field variables into “inside” 
($\lbrace\phi\rbrace_{A}$) and “outside” ($\lbrace\phi\rbrace_{B}$) parts, then one can rewrite the
ground state wave function as
\begin{eqnarray}\label{GroundSwavesplit} 
\Psi_0 \propto \exp\{-(\lbrace\phi\rbrace_A~ \lbrace\phi\rbrace_B)
\begin{pmatrix}
  \Gamma_{AA} & \Gamma_{AB} \\
  \Gamma_{BA} & \Gamma_{BB} 
 \end{pmatrix}
\begin{pmatrix}
  \lbrace\phi\rbrace_A \\
  \lbrace\phi\rbrace_B
 \end{pmatrix}
		\}~,
\end{eqnarray}
where $\Gamma_{\oplus \otimes}$ ($\oplus =\lbrace A,B\rbrace$ and $\otimes =\lbrace A,B\rbrace$)
 denotes the kernel matrix restricted to the inside or the outside. 

For the fields $\lbrace\phi^{1,2}\rbrace_{A}$ which are defined in the inside region, the ground state density
 matrix $\rho_A(\lbrace\phi^{1}\rbrace _{A},\lbrace\phi^{2}\rbrace_{A})$, is given by 
\begin{eqnarray}
\begin{split}
\rho_A(\lbrace\phi^1\rbrace_A; \lbrace\phi^2\rbrace_A) &\propto (\text{det}(\Gamma_{AA})^{-1})^{\frac{1}{2}}
	\exp\lbrace-{1\over 2}\times\\&(\lbrace\phi^1\rbrace_A~ \lbrace\phi^2\rbrace_A)
\begin{pmatrix}
  \mathcal{A} & 2\mathcal{B} \\
  2\mathcal{B} & \mathcal{A} 
 \end{pmatrix}
\begin{pmatrix}
  \lbrace\phi^1\rbrace_A \\
  \lbrace\phi^2\rbrace_A
 \end{pmatrix} \rbrace~,
\label{gaussdm}
\end{split}
\end{eqnarray}
where
\begin{eqnarray}
\mathcal{A} &=& 2(\Gamma_{AA} -{1 \over 2} \Gamma_{AB} (\Gamma_{BB})^{-1} \Gamma_{BA});\cr
\mathcal{B} &=& -{1\over2} \Gamma_{AB} (\Gamma_{BB})^{-1} \Gamma_{BA}~.
\label{ABdm}
\end{eqnarray}
From now on one can follow two different methods to get the entanglement entropy: one is based on direct diagonalization of the above 
reduced density matrix and the other based on using replica trick. For later use we will summarize the results for both of them. Using appropriate 
transformations \cite{Bombelli1986} one can write  the reduced density matrix as 
\begin{eqnarray}\label{gdstdm2}
\begin{split}
&\rho_A(\lbrace\phi^1\rbrace_A; \lbrace\phi^2\rbrace_A)=\prod_i \frac{1}{\sqrt{\pi}}\times \\  &\exp\lbrace -\frac{1}{2}(\phi_n^1\phi^{1n}+\phi_n^2\phi^{2n})-\frac{1}{4}E_i
(\phi^1-\phi^2)_n(\phi^1-\phi^2)^n\rbrace
\end{split}
\end{eqnarray}
where $E_i$'s are the eigenvalues of the matrix $\Lambda$ with the following simple form
\begin{equation}
\Lambda \equiv -(\Gamma^{-1})_{AB}~\Gamma_{BA}~.
\label{newker}
\end{equation}
The interesting point about the equation (\ref{gdstdm2}) is that it has the form of the reduced density matrix of two body  harmonic oscillator. In other words for the ground state
of coupled harmonic oscillator the problem of calculating the entanglement entropy can be reduced to the problem of 
calculating the entanglement entropy of two coupled harmonic oscillators. One can then show that the entropy can be expressed in terms of the eigenvalues $E_i$ of $\Lambda$
 as \cite{Bombelli1986}:
\begin{equation}\label{entsum}
S =\sum_{i} \left[\log \frac{\sqrt{E_i}}{2}+\sqrt{1+E_i} \log(\frac{1}{\sqrt{E_i}}+\sqrt{1+\frac{1}{E_i}})\right]~.
\end{equation}
It is worth mentioning that having larger coupling between two oscillators  leads to larger $E$ and consequently larger entanglement entropy.

The second method which is also useful to get  the R\'enyi entropy is based on Replica trick.
Using Eq. (\ref{gaussdm}), and rescaling the reduced density matrix one can calculate ${\rm \tr}\rho_A^n$ 
 and ultimatley the entropy \cite{Callan1994} as 
following sum 
\begin{equation}
S =\lim_{n\rightarrow 1} \frac{1}{1-n}\log(\rm \tr \rho_A^n)=-\sum_{i=1}^{l}\{{\rm ln}(1-\xi_i) +
{\xi_i\over 1-\xi_i} {\rm ln}\xi_i\}~,
\label{eigsum}
\end{equation}
where $\xi_i$ is related to the eigenvalue of the matrix $\mathcal{C}=-2\mathcal{A}^{-1}\mathcal{B}$ by $\mathcal{C}_i=\frac{2\xi_i}{1+\xi_i^2}$ .
 
It is also useful to consider the matrix $\Lambda=(1-\mathcal{P})^{-1}\mathcal{P}$ where
 $\mathcal{P} \equiv \Gamma_{AA}^{-1}\Gamma_{AB} \Gamma_{BB}^{-1}\Gamma_{BA}$ which has also the simple form (\ref{newker})
and write Eq. (\ref{eigsum}) in terms of eigenvalues of the matrix $\Lambda$ as (\ref{entsum}). 
 The eigenvalues $E_i$  of the matrix $\Lambda$ are positive and related to $\xi_i$ by
\begin{equation}\label{xi and E}
\xi_i = \frac{\sqrt{1+E_i}-1}{\sqrt{1+E_i}+1}~.
\end{equation}

It is also straightforward to write the R\'enyi entropy $S_n$ in term of $\xi_i$ as: 
\begin{eqnarray} \label{Renyi} 
S_n = \frac{1}{n-1}\sum_i \left( \log(1-\xi_i^n) -n\log(1-\xi_i) \right)~.
\end{eqnarray}
In order to compute the entanglement entropy obtained by tracing over the fields 
in the region $A$ for a given problem, one should find the eigenvalues of the 
matrix $\Lambda$. For a given hamiltonian $\mathcal{H}$ one can easily find the operators $K$ 
and consequently $\Gamma$ and $\Gamma^{-1}$.  In the continuum 
limit, the operator $\Lambda$ is obtained after integration over the oscillators in the region $B$ as
\begin{equation}
\Lambda(x,y)=-\int_B dz \Gamma^{-1}(x,z)\Gamma(z,y).
\label{Lambdacalc}
\end{equation}  
The eigenvalue problem to be solved is then
\begin{equation}\label{eigvalproblem}
\int dy \Lambda(x,y)\psi(y)=E\psi(x),
\end{equation} 
where $\psi(x)$ is an eigenfunction with eigenvalue $E$.

It is worth mentioning that for the general Hamiltonian Eq. ($\ref{harmonicOsc}$), 
one can calculate the two point correlators $X_A=\rm \tr(\rho_A \phi_i \phi_j)$ and 
$P_A=\rm \tr(\rho_A \pi_i \pi_j)$ using the $K$ matrix by

\begin{eqnarray}\label{X_A P_A}
\frac{1}{2}K^{-1/2}=
\begin{pmatrix}
  X_{A} & X_{AB} \\
  X^{ T}_{AB} & X_{B}
 \end{pmatrix}, \hspace{0.2cm} \frac{1}{2}K^{1/2}=
\begin{pmatrix}
  P_{A} & P_{AB} \\
  P^{ T}_{AB} & P_{B} 
 \end{pmatrix}~,
\end{eqnarray}
Then one can define  matrix $C = \sqrt{X_AP_A}$, which has the eigenvalues \cite{Casini2009},
\begin{equation} \label{XPspectrum} 
\nu_i=\coth(-\log(\frac{\sqrt{1+E_i}-1}{\sqrt{1+E_i}+1})/2),
\end{equation} 
where $\nu_i$ are the eigenvalues of $C$. With respect to the new operators the entropy is given by
\begin{eqnarray} \label{Entropy from corr} 
S &=\rm \tr\left[(C+\frac{1}{2})\log(C+\frac{1}{2})-(C-\frac{1}{2})\log(C-\frac{1}{2})\right]\nonumber \\
& = \sum_{i=1}^l \left[(\nu_i+\frac{1}{2})\log(\nu_i+\frac{1}{2})-(\nu_i-\frac{1}{2})\log(\nu_i-\frac{1}{2})\right]~.
\end{eqnarray} 
We also have
\begin{eqnarray} \label{Renyi from corr} 
 S_n &=& \frac{1}{n-1}\rm \tr\left[ \log\left((C+\frac{1}{2})^n-(C-\frac{1}{2})^n\right) \right]\nonumber \\
 &=& \frac{1}{n-1}\sum_{i=1}^l \left[ \log\left((\nu_i+\frac{1}{2})^n-(\nu_i-\frac{1}{2})^n\right)\right],
\end{eqnarray}
where $l$ is the size of the subsystem $A$.
In this formulation  we need only the correlators inside the region $A$, to calculate $S$ and $S_n$.

In order to clarify the hamiltonian approach in the continuum limit, we briefly review the procedure followed in \cite{Callan1994} to find 
an approximate analytical solution for the harmonic oscillator system with short-range 
interaction. This method is introduced in order  to determine $E$ and also $S$
 by using eigenvalue problem Eq. (\ref{eigvalproblem}). They considered a one dimensional 
coupled harmonic oscillator with mass $M$, confined to the region $-L<x<L$ and the subsystem 
is taken to be half of a finite system. 

To calculate the eigenvalue $E$, for  system of 
harmonic oscillators with short-range interactions, it is better to first consider a system with 
infinite size $L\rightarrow \infty$. At this limit the kernels $\Gamma^{\pm 1}$ needed 
to construct $E$ have the following forms:
\begin{eqnarray}
\begin{split}
&\Gamma(x,y) = M K_1(M(x-y))/(\pi(x-y))~,\\ &\Gamma ^{-1}(x,y)=K_0(M(x-y))/\pi~.
\end{split}
\end{eqnarray}
where $M$ is the mass term. In the $M\rightarrow 0 $ limit it is easy to show that $\psi=\exp(\imath \omega \ln x)$ is an eigensolution of the Eq. (\ref{eigvalproblem}) with eigenvalue
\begin{equation}\label{energy callan}
E ={\rm sinh}^{-2}(\pi\omega)~.
\end{equation}
To discretize the spectrum and calculate the entropy, one needs to 
impose Dirichlet boundary conditions at some large $x=L$ and further Dirichlet condition at some small $x=\epsilon$. 
The eigenvalues and eigenvectors are then: 
\begin{equation}\label{Callaneigenvector}
\psi(x) = \sin(\omega(E)\ln(x/\epsilon)), \hspace{.2cm}\omega(E_i) \ln(L/\epsilon)=\pi i~.
\end{equation}  

It is  useful to note that the density of states per unit $\omega$ interval is constant. 
Now one can rewrite the continuum limit of the R\'enyi entropy Eq. (\ref{Renyi}) and the entanglement entropy 
 as an integral over $\omega$,
\begin{equation}\label{Renyi continuum}
S_n = \frac{\log L}{\pi (n-1)} \int_0^{\infty}d\omega (\log(1-\xi^n)-n\log(1-\xi)),
\end{equation} 
\begin{equation}\label{Renyi n 1 continuum}
S = \frac{\log L}{\pi } \int_0^{\infty}d\omega \left( \frac{\xi}{\xi-1}\log(\xi) - \log(1-\xi) \right),
\end{equation} 
where $\xi(\omega)$ is defined in the Eq. (\ref{xi and E}). 

As discussed before, conformal invariance implies universal properties for the entanglement entropy.
 The entanglement entropy and also the R\'eyni entropy for these models, diverge logarithmically with
 the subsystem size with prefactors proportional to $c$ and $c_n$, respectively. 

By using the Eq. (\ref{Renyi n 1 continuum}) and also Eqs. (\ref{xi and E}) and (\ref{energy callan}) 
one can find the entanglement entropy $S$ for the harmonic oscillator problem, giving the result $S=\frac{1}{6}\log(L/\epsilon)$
which is consistent with $c=1$. In addition using (\ref{Renyi continuum}) one 
can also find the R\'enyi entropy $S_n=\frac{1}{12}\left(1+\frac{1}{n}\right)\ln(L/\epsilon)$
consistent with the CFT predictions \cite{Calabrese2004}.   

Next we  consider short-range harmonic oscillator with infinite size and sub-system with length $l$. This kind of configuration is completely different 
from the Ref. \cite{Callan1994}. We should remember that the Eqs. (\ref{energy callan}) and 
(\ref{Callaneigenvector}) are no longer true in this configuration. We proceed
 with the Eq. (\ref{Lambdacalc}). To evaluate this integral we must 
consider $B\in\left(-\infty<z<0\right)\cup \left( l<z<\infty\right)$ as the complement 
of the sub-region $l$. The matrix $\Lambda$ becomes:
\begin{eqnarray} \label{lambda matrices SRHO small l}
 \Lambda(x,y)&=&-\frac{1}{\pi^2}\int_{B}\frac{\ln(x+z)}{(z+y)^2}dz \nonumber\\
 &=&\frac{1}{\pi^2}\bigg\lbrace\frac{\left(l-x\right) \log(l-x)-\left(l-y\right) \log(l-y)}{(l-y) (y-x)}\nonumber \\
 &-&\frac{x \log(x)-y \log(y)}{(y-x) y}\bigg\rbrace.
\end{eqnarray} 
Therefore, according to Eq. (\ref{eigvalproblem}), the eigenvalues $E_i$ and the 
corresponding eigenfunctions $\psi_i(x)$ can be obtained by diagonalizing $\Lambda$ 
matrix. Unfortunately we were unable to find $E_i$ analytically. One can numerically evaluate $E_i$ and $\psi_i$ using direct diagonalization 
of the matrix $\Lambda$, then try to guess the formula for eigenvalues and eigenfunctions.
 We shall come to this problem in the next section by means of numerical calculations.

We are now ready to speak more about the LRHO with $\alpha<2$. To determine $E$ and $\psi$
 for LRHO, we calculated first the matrices $\Gamma=K^{1/2}$ and $\Gamma^{-1}=K^{-1/2}$.
 The continuum limit of the matrices $\Gamma$ and $\Gamma^{-1}$ has the following forms:

\begin{eqnarray} \label{W matrices}
\begin{split}
 \Gamma^{\pm 1}(x,y)&=\frac{1}{2\pi}\int_{-\infty}^{\infty} dk(|k|^{\alpha}+M^{\alpha})^{\pm 1/2}e^{ik.(x-y)}\\   
  &= \frac{1}{2\Gamma[\mp \alpha/2]}\frac{1}{|r|^{1\pm \alpha/2}}\times \\
  &H_{3,2}^{1,2}\left( \left(M|r| \right)^\alpha \middle|
 \begin{array}{c}
      (1,1)(\mp \frac{\alpha}{2},1)(\mp \frac{\alpha}{4},\frac{\alpha}{2})\\
      (\mp \frac{\alpha}{2},1)(\mp \frac{\alpha}{4},\frac{\alpha}{2})
  \end{array}\right)\\
 &=\frac{1}{2\Gamma[\mp \alpha/2]\cos(\frac{\pi\alpha}{4})}\frac{1}{|r| ^{1\pm \alpha/2}}+\mathcal{O}(M^{\alpha})~,
\end{split}
\end{eqnarray}  
where $r=x-y$ and $H_{3,2}^{1,2}$ is the Fox H-Function. 
Then in a similar way as in Eq. (\ref{Lambdacalc}), we found the matrix $\Lambda$ by 
multiplying $\Gamma$ and $\Gamma^{-1}$ in the complement region $B\in\left(-\infty<z<0\right)\cup \left( l<z<\infty\right)$
 as 

\begin{eqnarray}\label{lambdamatrix}
\Lambda(x,y) = 
    \mathfrak{A}\left[\frac{2 \left(\left(\frac{l-x}{l-y}\right)^{\alpha/2}-\left(\frac{x}{y}\right)^{\alpha/2}\right)}{\alpha (x-y)}\right] \hspace{.1cm}(\alpha < 2)
\end{eqnarray}
where  $\mathfrak{A}=\frac{1}{4\Gamma[-\alpha/2]\Gamma[\alpha/2]\cos^2(\frac{\pi\alpha}{4})}$. 
Unfortunately, the exact solution of the eigenvalues and the corresponding eigenfunctions for
 Eq. (\ref{lambdamatrix}) are not known and remain an open problem. It is nonetheless both 
possible and interesting to investigate the properties of the $E$ and $\psi$, for LRHO problem, 
numerically. In the next section we will discuss our numerical findings.

It is worth mentioning that, the Eq. (\ref{lambdamatrix}) is only true for an infinitely 
large system compared to the sub-system size. However one can also study 
LRHO problem in the presence of boundary but since boundary of the finite system breaks 
the translational invariance, we have not been able to find $\Gamma^{\pm 1}$ explicitly because
 we are not allowed to use Fourier transform for the finite systems. Therefore we studied 
this case just numerically, which we will present the results in the next sections.

As explained above, we can find the eigenvalues $E$ and the corresponding eigenvectors $\psi(x)$ for 
a given matrix $K$, and  we can also study the scaling behaviors of the entanglement entropy $S$ and the R\'enyi entropy $S_n$. 
 
\begin{figure}[htp]
\begin{center}
\includegraphics[width=9cm,clip]{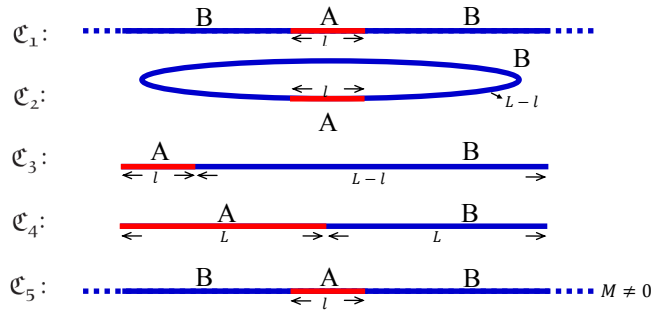}
\end{center}
\caption{(Color online) Different configuration of systems and subsystems.}
\label{Figure:1}
\end{figure}

In the next section we will speak more about our results but here we will discuss about different configurations 
for the system and also subsystem that  we have used in our study. 
In this work we consider five main kinds of configurations depicted in Fig. (\ref{Figure:1}) for system and subsystem. In the massless case:
\begin{itemize}\item[$\mathfrak{C_1}$:] System is very large and $A$ is a small sub-system with length $l$. 
\item[$\mathfrak{C_2}$:] System with periodic boundary condition has finite size $L$, and $A$ is a sub-system with length $l$.
\item[$\mathfrak{C_3}$:] System with size $L$ has boundary and is divided to two adjacent parts. The 
first part is a sub-system with length $l < L$ and the second part is the complement with size $L-l$.
\item[$\mathfrak{C_4}$:] System with size $2L$ has boundary and is divided to two adjacent equal intervals 
with length $l=L$ where one of them is the sub-system.

\end{itemize}  
In the massive case:
\begin{itemize}
\item[$\mathfrak{C_5}$:] System is very large and $A$ is a sub-system with length $l$.
\end{itemize}  
\subsection{Numerical evaluation}\label{sec3}

We now numerically evaluate the von Neumann entanglement entropy $S$ and the R\'enyi 
entropy $S_n$ for LRHO problem in different cases ($\mathfrak{C_i}$,$i=1\dots5$), 
by using Eqs. (\ref{entsum}) and (\ref{Renyi}) or equivalently Eqs. (\ref{Entropy from corr})
 and (\ref{Renyi from corr}), which was first studied in \cite{Nezhadhaghighi2012}. In this respect, 
we follow the method explained in the last section. We will measure the eigenvalues $E_i$ and 
the eigenfunctions $\psi_i(x)$ in Eq. (\ref{eigvalproblem}) numerically and then we introduce 
an expression for $E$ and $\psi$, which matches to the numerical simulations. Our motivation 
to study these quantities with full detail is related to our interest in better understanding the
operator (\ref{lambdamatrix}) which its eigenvalues provide the entanglement entropy. We should here stress that we calculate the
entanglement entropy using the numerical $\Lambda$ matrix and not by discretizing the operator (\ref{lambdamatrix}). However 
we will confirm that these two operators are very close to each other if we consider large systems and consequently can approximate each other.

In order to calculate $E$ and $\psi$, we first need to construct the matrix $\Lambda$ for 
a given $K$ matrix. 
Numerically one can find the matrix $\Lambda\equiv -\Gamma^{-1}_{+-}\Gamma_{-+}$  
by multiplying $\Gamma^{-1}$ and $\Gamma$, where $\Gamma=K^{1/2}$ and $\Gamma^{-1}=K^{-1/2}$. 
For example we applied this method to the LRHO with very large system size and small sub-region
 $l$. There is a very good agreement between numerical $\Lambda$ and the matrix $\Lambda(x,y)$ 
coming from Eqs. (\ref{lambda matrices SRHO small l}) and (\ref{lambdamatrix}), when the distances 
are more than four lattice sizes. 
\begin{figure}[htp]
\begin{center}
\includegraphics[width=9cm,clip]{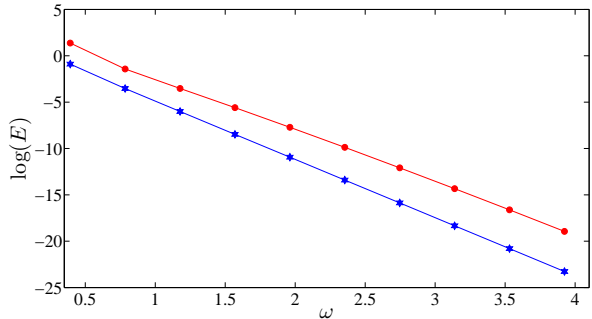}
\end{center}
\caption{(Color online)  The eigenvalues $\log(E_i)$ versus $\omega_i$ 
for HO with short range interaction with the configuration
 $\mathfrak{C_4}$. The blue stars correspond to $E = 1/\sinh(\pi \omega)^2$, where
 $\omega=n\pi/\log(L)$.}
\label{Figure:2}
\end{figure}

To obtain a better understanding of the long-range harmonic oscillator problem, we studied first 
the eigenvalues $E_i$ and the eigenfunctions $\psi_i$ of the short-range harmonic oscillator. 
We considered a system with size $2L$ and the subsystem is taken to be half of the system 
size ($\mathfrak{C_4}$). Then using the numerical methods, we diagonalized the matrix $\Lambda$ 
to find $E_i$ and $\psi_i$. In Fig. (\ref{Figure:2}) we sketched  logarithm of the eigenvalues
 $E_i$ with respect to $\omega(E_i)$. As can be seen, the result obtained from Eq. (\ref{energy callan})
 has similar asymptotic behavior as numerical simulations. In addition the eigenvectors $\psi(x)$ for the first
 and second largest eigenvalues, $E_1$ and $E_2$  verify the behavior 
predicted in Eq. (\ref{Callaneigenvector}). We have also calculated the prefactor $c$ numerically and our
result is consistent with the theoretical prediction. The numerical results of entanglement
entropy for LRHO ($\alpha<2$) for the systems with boundary e.g. $\mathfrak{C_3}$ and $\mathfrak{C_4}$, 
are summarized in the next sections. 

\begin{figure}[htp]
\begin{center}
\includegraphics[width=9cm,clip]{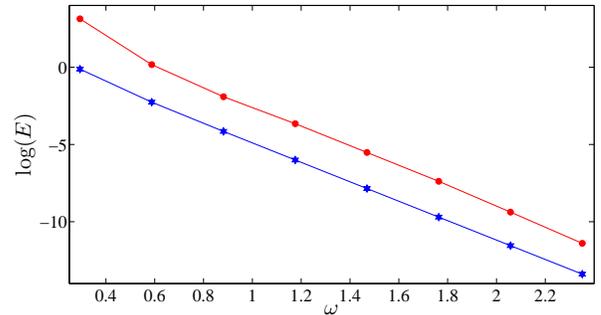}

\end{center}
\caption{(Color online)  The eigenvalues $\log(E_i)$ versus $\omega_i$ for HO with short range 
interaction with the configuration $\mathfrak{C_1}$. The blue stars correspond to
$E = 1/\sinh(\pi \omega)^2$, where $\omega=n\pi/2(\log(l)+\zeta)$ and $\zeta = 1.3$.  
}
\label{Figure:3}
\end{figure}

Next we discuss the case, where the subsystem is very small with length $l$ and the system is 
very large ($\mathfrak{C_1}$). For this configuration, as a first step, we have studied the 
properties of $E_i$ and $\psi_i$ for harmonic oscillator problem with short-range interaction 
by a direct diagonalization of the matrix $\Lambda$. 
Numerical results are shown in Fig. (\ref{Figure:3}). It is interesting to note that, when we choose 
$\omega(E_i)=\pi i/2(\log(l)+\zeta)$ ($\zeta = 1.3$),
apart from a constant, which it appears ubiquitously in this kind of studies \cite{Peschel2003},
 the behavior of the eigenvalues $E_i$ are in very good agreement 
with $E(\omega) = 1/\sinh^2(\pi \omega)$  (see Fig . (\ref{Figure:2})). Let us remark that 
$\omega(E)$ for the configuration $\mathfrak{C_1}$, differs from Ref. \cite{Callan1994} by a factor two 
and a constant $\zeta$. 
We studied the scaling of $S$ versus the logarithm of the sub-system size, $\log l$, and 
compared with Eq. (\ref{scalingEsmall_l}). Our result agrees  with $c = 1$.

\begin{figure}[htp]
\begin{center}
\includegraphics[width=9cm,clip]{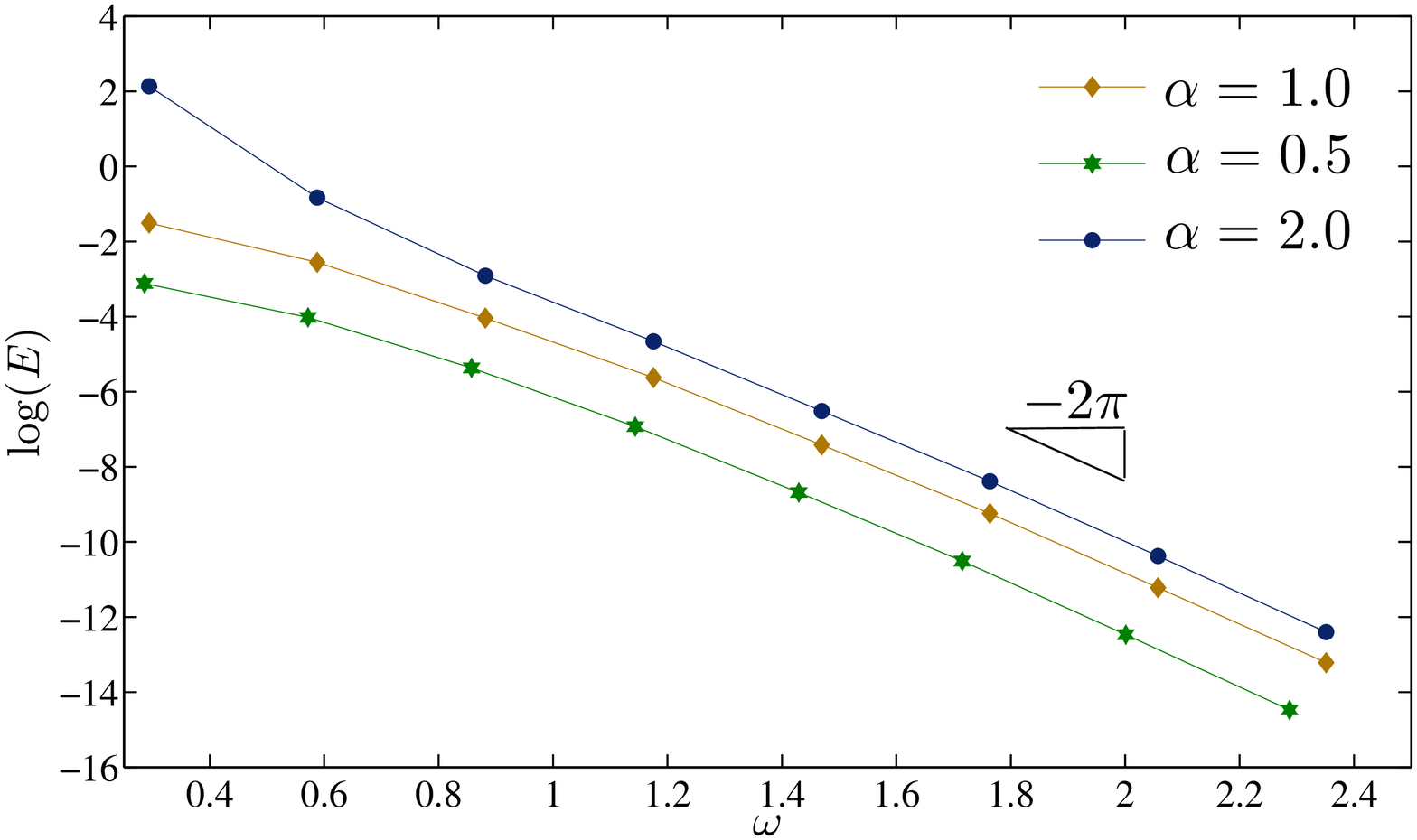}
\end{center}
\caption{(Color online) The eigenvalues $\log(E_i)$ versus $\omega_i$ for 
LRHO with the configuration $\mathfrak{C_1}$ and different $\alpha$'s. 
The small eigenvalues (large $\omega_i$) are independent of $\alpha$ and  $\log(E_i)$ is linearly dependent to $\omega_i$ by scaling factor $-2\pi$.}
\label{Figure:4}
\end{figure}

The next step is to analyze the eigenvalues $E$ of the Eq. (\ref{lambdamatrix}) 
for LRHO with $\alpha<2$. As we remarked before, if we consider very small sub-region 
of LRHO with $\alpha=2$ and very large system size ($\mathfrak{C_1}$), we expect 
$E_i \sim \sinh^{-2}(\pi \omega_i)$ and $\omega(E_i) = i\pi/2(\log(l)+\zeta)$. For 
other values of $\alpha$, the eigenvalues behavior can be seen in the Fig. (\ref{Figure:4}), 
where we compared $\log(E)$ \textit{vs.} $ \omega$ for various $\alpha$'s. Let us first address 
the behavior of small eigenvalues $E_i$ (large $i$, i.e. $i>3$). Our results show that the 
small eigenvalues are independent of $\alpha$ and  $\log(E_i)$ is linearly dependent to 
$\omega_i$ by scaling factor $-2\pi$. Then one can get the asymptotic behavior 
$E_i \propto e^{-2\pi \omega_i}$ for $i>3$ and from our previous knowledge about $E_i$ for 
$\alpha=2$, one can conjecture the simple behavior $E_i\propto\sinh^{-2}(\pi \omega_i)$. 
In our numerical simulations we used   $\omega_i=i\pi/2(\log(l)+\zeta)$, where $\zeta$ 
is a $\alpha$ dependent parameter ( $\zeta \in \left[1.0,2.0\right]$), to get the best fit 
to numerical data. We may use this behavior and guess the asymptotic expression for the eigenvalue $E$ as
\begin{eqnarray}\label{energy LRpSR}
E(\omega) =
   \frac{a(\alpha)}{\sinh^2(\pi \omega)+b(\alpha)}.
\end{eqnarray}
The best 
fit parameters to our numerical data were $a(\alpha) = \frac{\alpha}{2}\sin^2 (\frac{\pi \alpha}{4})$ and 
$b(\alpha) = 0.12 \alpha +0.19 \alpha^2-0.20\alpha^3+0.04\alpha^4$ . The value of $b(\alpha)$ is zero at $\alpha=0$ 
and $\alpha=2$ and it has a maximum at $\alpha=1$. 

\begin{figure}[htp]
\begin{center}
\includegraphics[width=9cm,clip]{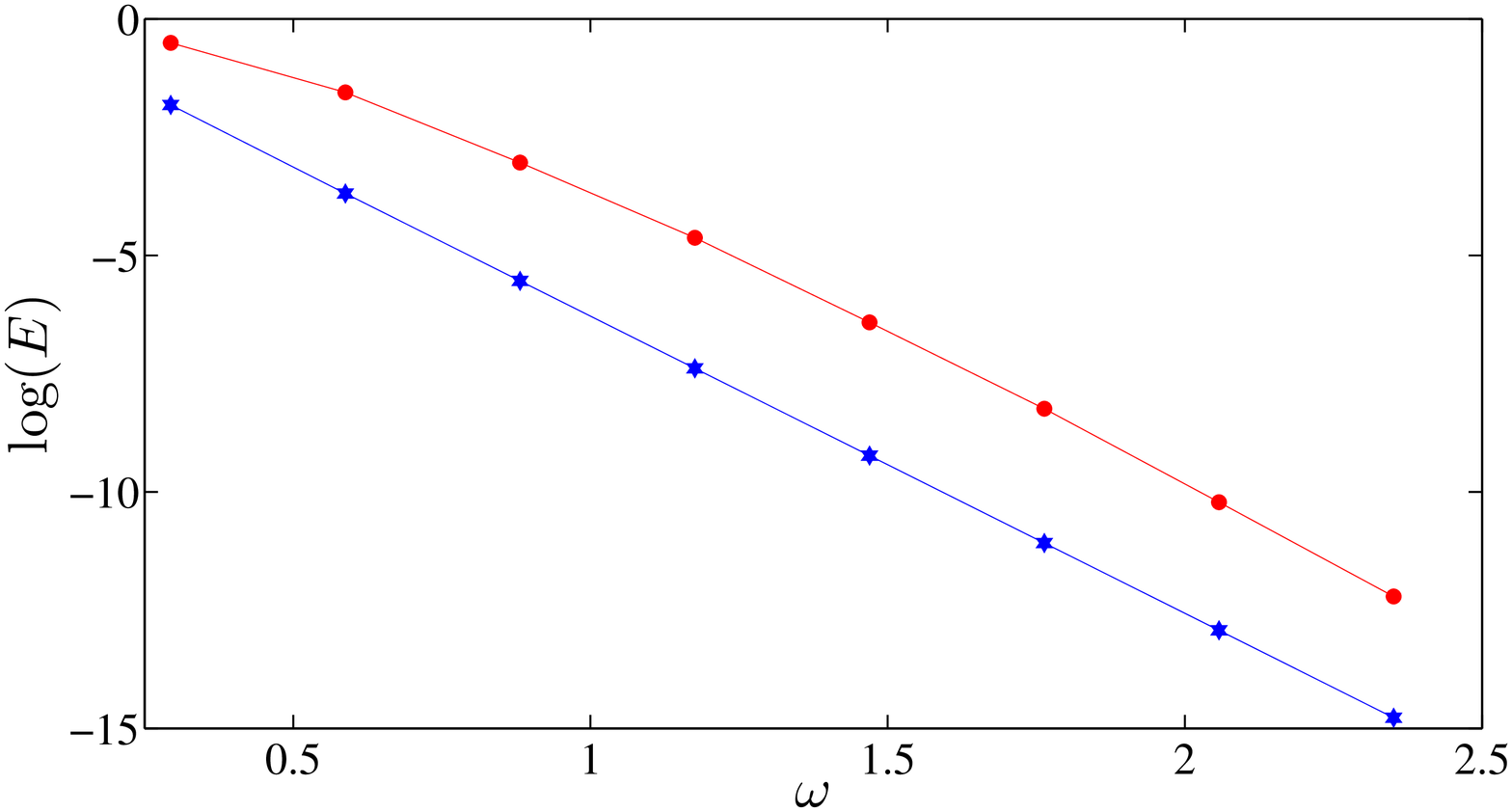}
\includegraphics[width=9cm,clip]{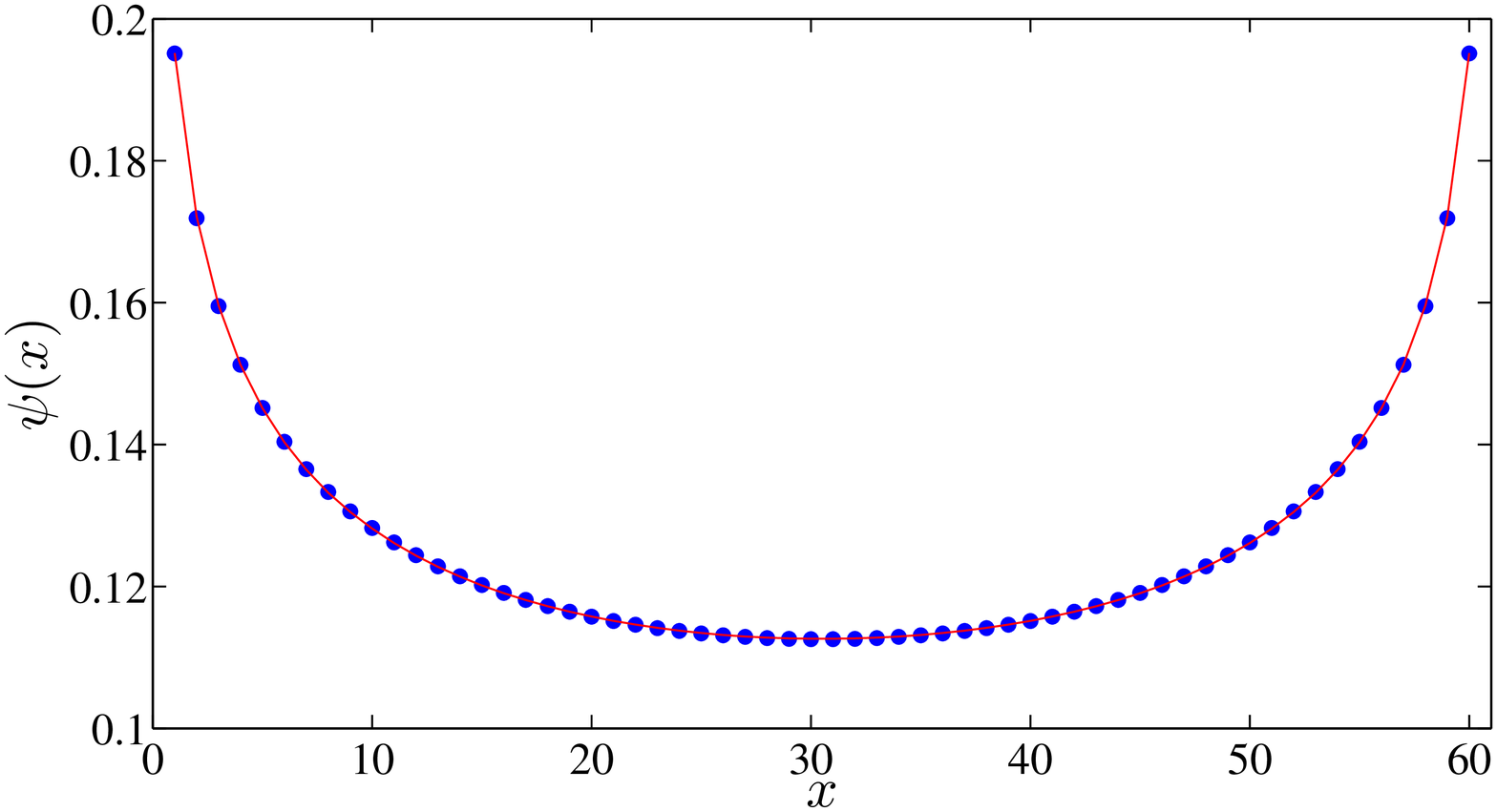}
\includegraphics[width=9cm,clip]{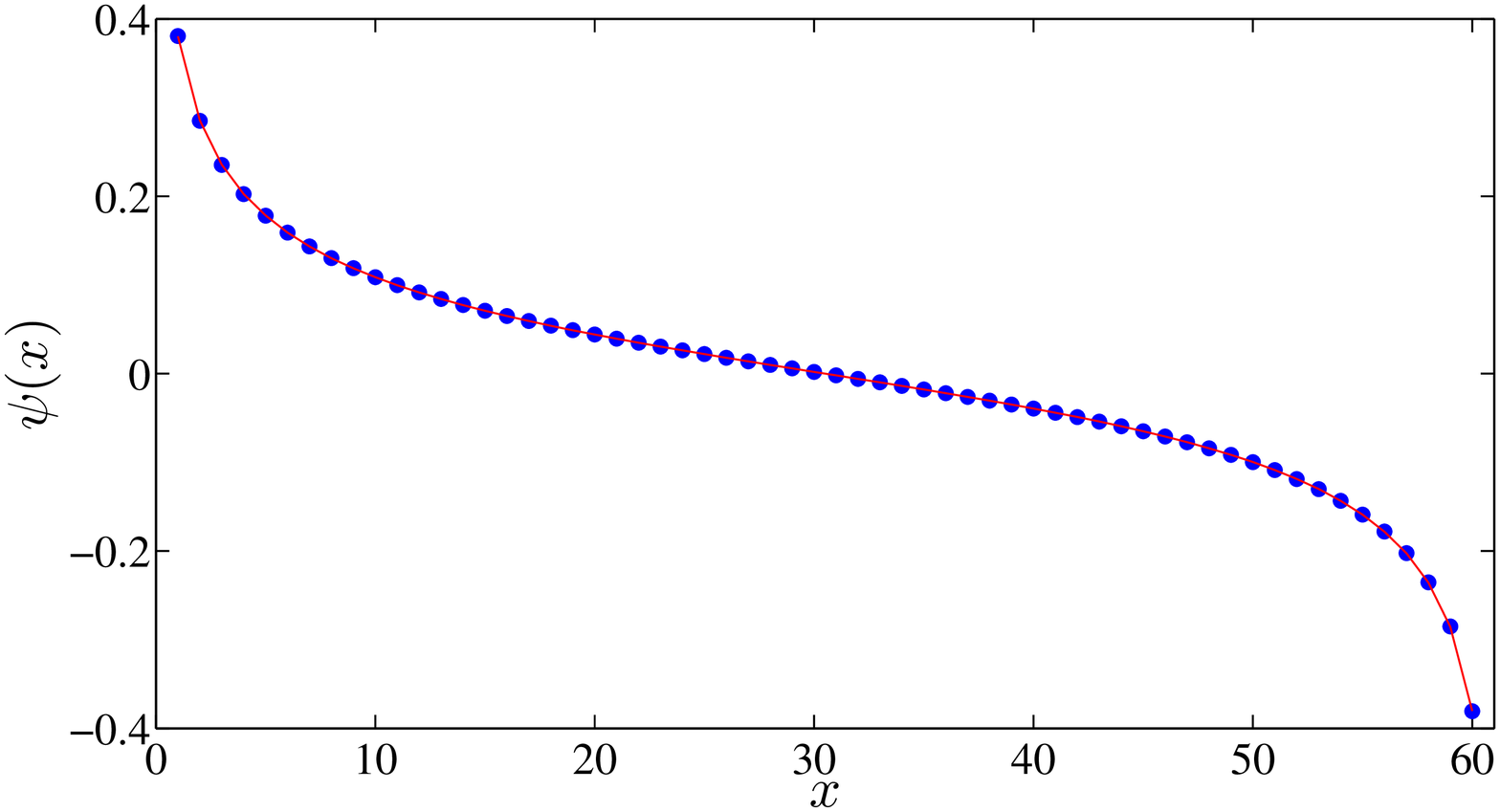}
\end{center}
 
\caption{(Color online) 
Top: The eigenvalues $\log(E_i)$ versus $\omega_i$ for LRHO ($\alpha=1$) 
with the configuration $\mathfrak{C_1}$. The blue stars correspond to
$E = \frac{a(1)}{\sinh^2(\pi \omega)+b(1)}$, where $\omega=n\pi/2(\log(l)+\zeta)$
 ($\zeta = 1.26$) and $a(1) = 0.25$ and also $b(1) = 0.14$. 
Middle: The eigenfunction $\psi_1(x)$ corresponds to the first eigenvalue $E_1$. 
Bottom: The eigenfunction $\psi_2(x)$ corresponds to $E_2$. Solid red lines correspond 
to normalized form of equation (\ref{trialeigenfunction}) ($b = -0.26,-0.34$ for $n=1,2$ respectively).}
\label{Figure:5}
\end{figure}

Next, we studied the eigenvector $\psi_i(x)$ of the matrix $\Lambda$ for LRHO numerically. 
By diagonalizing $\Lambda$ we can also find the eigenvalues $E_i$. The eigenvector $\psi_i$ 
can then be computed for each $E_i$ by $\Lambda \psi_i = E_i\psi_i$. A typical example is 
shown in Fig.  (\ref{Figure:5}), where one can see that the eigenfunctions
 $\psi_i(x)$ are symmetric around $x=l/2$ for odd $i$, and antisymmetric for even $i$. We 
found that the best fit to the eigenvectors $\psi_i(x)$ is
\begin{eqnarray}\label{trialeigenfunction}
\psi_i(x)&=&\frac{1}{\mathcal{N}}\lbrace \left( x^{\imath\omega_i+b}+x^{-\imath\omega_i+b}\right) \nonumber \\
&-& (-1)^i \left((l-x)^{\imath\omega_i+b}+(l-x)^{-\imath\omega_i+b}\right) \rbrace;
\end{eqnarray}
where $\mathcal{N}$ is the normalization coefficient and $b\in \left[ -1,0\right]$ 
is the free parameter to get the best fit to the numerical data. In Eq. (\ref{trialeigenfunction}) 
we used  $\omega_i=\frac{i \pi}{2(\log l+\zeta)}$. The values of the free parameters $b$ and $\zeta$ in 
general depend on $\alpha$. In Fig. (\ref{Figure:5}) the behavior of the $\psi$ for LRHO problem
 with $\alpha=1$ and also the best fit to the Eq. (\ref{trialeigenfunction}), are shown, as a function of $x$.

As argued before, we studied the eigenvalue problem Eq. (\ref{eigvalproblem}) 
for LRHO, in order to find the eigenfunction $\psi_i(x)$ and corresponding eigenvalue $E_i$. 
The von Neumann entanglement entropy $S$ and the R\'enyi entropy $S_n$ can be obtained as  functions of $E$ 
(see Eq. (\ref{Renyi})). It is possible to find $S$ and also $S_n$ using Eq. (\ref{Entropy from corr} ) 
and Eq. (\ref{Renyi from corr}), respectively. 

Finally, we discuss the goodness of Eq. (\ref{energy LRpSR}). For arbitrary values of the long-range interaction 
$\alpha$, the von Neumann entanglement entropy $S$ and the R\'enyi entropy $S_n$ can in practice be
 obtained by $(i)$ evaluating Eq. (\ref{HOLR}) numerically for system with total size $L$ and compute
 $X_A$ and $P_A$ from $K$ matrix (see Eq. (\ref{X_A P_A})), $(ii)$ diagonalizing $C$ to obtain $\nu_i$, 
and $(iii)$ evaluating (\ref{Entropy from corr}) and (\ref{Renyi from corr}), where $l$ is the number of lattice sites in the subsystem $A$. 

\begin{figure}[htp]
\begin{center}
\includegraphics[width=9cm,clip]{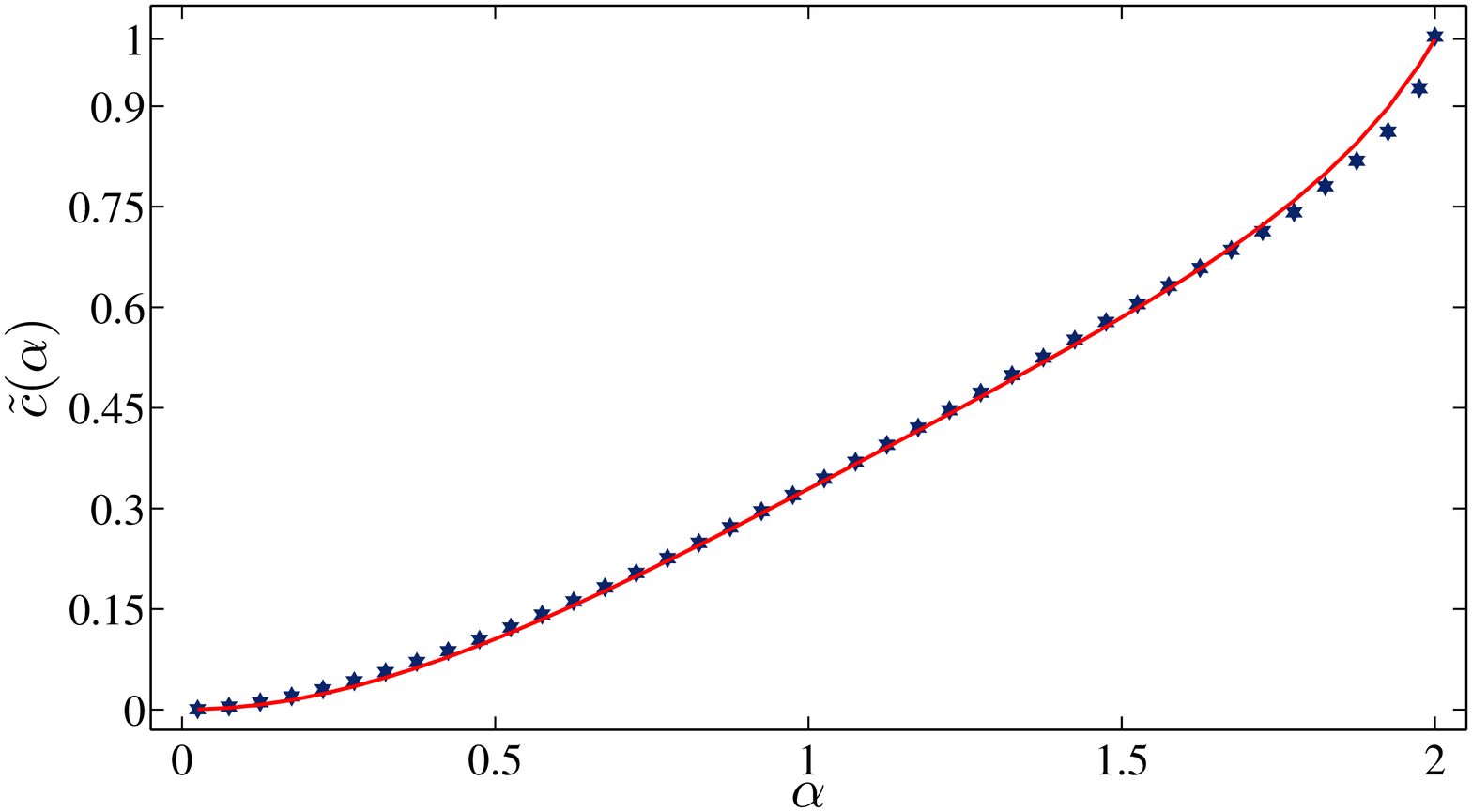}
\includegraphics[width=9cm,clip]{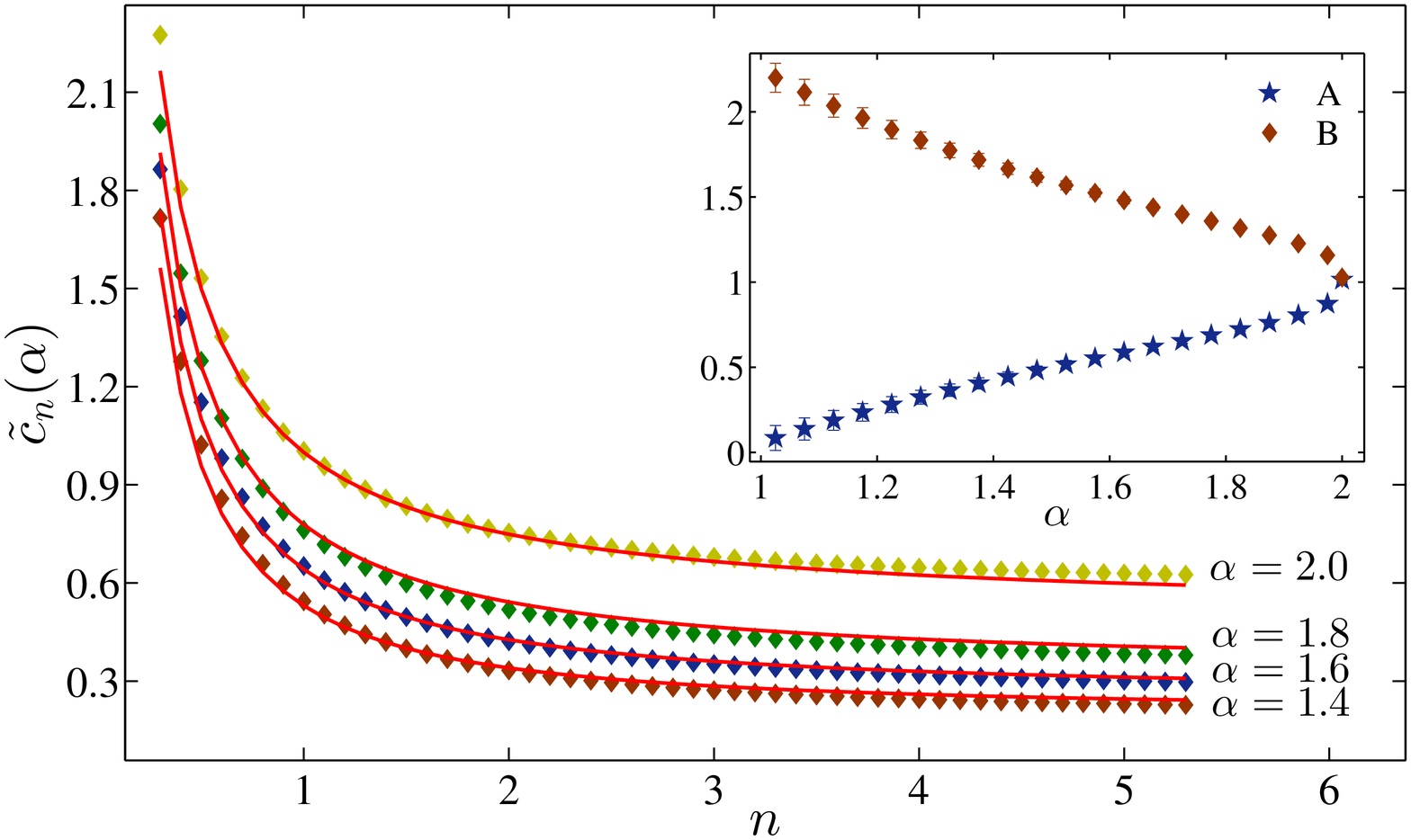}
\end{center}
\caption{(Color online) Top: The prefactor $\tilde{c}(\alpha)$ for discrete system with size $L=6000$ with 
the configuration $\mathfrak{C_1}$. The prefactor is measured using the scaling relation
 $S$ with $\log l$ in the region $0<l<L/100$. The red line represents the same quantity coming 
from the continuum limit approximation. Bottom: $\tilde{c}_n(\alpha)$ vs. $n$ for different $\alpha$'s 
(from top to bottom: $\alpha = 2.0, 1.8, 1.6, 1.4$. The red lines come from the continuum limit 
approximation. Inset: $A$ and $B$ coefficients vs. $\alpha$.}
\label{Figure:6}
\end{figure}

We observe that, in the LRHO problem the entanglement entropy and the R\'enyi entropy increase
 logarithmically  with the sub-system size as 
\begin{equation}
S\sim \frac{\tilde{c}(\alpha)}{3}\log l~,
\end{equation}
and
\begin{equation}
S_n\sim \frac{\tilde{c}_n(\alpha)}{3}\log l~,
\end{equation} 
respectively. By studying the scaling behavior of $S$ and also $S_n$ vs. $\log l$, 
one can find the scaling parameters $\tilde{c}(\alpha)$ and $\tilde{c}_n(\alpha)$.
 We display the resulting quantities for different values of $\alpha$ and $n$, in Fig. (\ref{Figure:6}). 

For arbitrary values of $\alpha$ and $n$, according to Eqs. (\ref{Renyi continuum}) and 
(\ref{Renyi n 1 continuum}) and also Eq. (\ref{energy LRpSR}), one can 
find the prefactors $\tilde{c}(\alpha)$ and $\tilde{c}_n(\alpha)$. We have depicted the 
results coming from these formulas in Fig. (\ref{Figure:6}), and we found perfect
agreement between our results, confirming the validity of the Eq. (\ref{energy LRpSR}).
There are some comments in order: the fact that the coefficient of the logarithm is an increasing function of $\alpha$ is somehow counter
intuitive because we know that for bigger $\alpha$'s the interaction get weaker by the distance faster than the smaller $\alpha$'s.  There are some ways to  roughly 
understand 
this result: from
mathematical point of view one might argue that the entanglement entropy is actually related to the eigenvalues of the matrix $\Lambda$
and those eigenvalues are smaller when we take smaller $\alpha$'s. This can be seen easily by looking to the equation (\ref{lambdamatrix}). These eigenvalues
are also the parameters that appear after mapping the many body harmonic oscillator to the two body case in equation (\ref{gdstdm2}). Stronger couplings between two 
oscillators leads to bigger $E$ and consequently bigger entanglement among them. The fact that after diagonalization we have smaller $E_i$'s for smaller $\alpha$'s
shows that although the interactions between oscillators far from each other is much stronger for smaller $\alpha$'s, that still does not guaranty bigger entanglement entropy. 
One might understand this phenomena as follows: based on the equation (\ref{HOLR}) in the range $0<\alpha<2$  one can see that $K(1)$, which is 
related to the nearest neighbor interaction, is 
an increasing function with respect to $\alpha$ but $K(n)$ with $n>1$ first increases with $\alpha$ and then decreases. It seems like the value of $E_i$
is mostly dependent on the value of the nearest neighbor interaction and follows the same trend. So although in some range of $\alpha$'s the next nearest neighbor interaction
for bigger $\alpha$ is smaller the  entanglement after considering the nearest neighbor interaction is bigger. This also explain qualitatively why we get an increasing 
function of $a(\alpha)$ in the equation (\ref{energy LRpSR}). This reasoning is consistent with the
area law observation in the massive case and also higher dimensions  that we are going to discuss later.

We now turn to determine the behavior of $\tilde{c}_n(\alpha)$  with respect to $n$.
 Interestingly, we find that the best fit to $\tilde{c}_n$ is
\begin{equation}
\tilde{c}_n(\alpha)=\frac{\tilde{c}(\alpha)}{2}(A(\alpha)+\frac{B(\alpha)}{n})~.
\end{equation}
The coefficients $A(\alpha)$ and $B(\alpha)$ are functions of $\alpha$ (see Fig. (\ref{Figure:6})), 
which indicates that LRHO is not conformally invariant(notice that by definition $A(\alpha)+B(\alpha)=2$). In conformal invariant systems 
$\tilde{c}_n = \frac{c}{2}(1+\frac{1}{n})$, where $c$ is the central
charge of the system. At this point it is worth mentioning that one can also calculate single copy entanglement introduced in \cite{Eisert2005}. Since this quantity is
equivalent to the R\'enyi entropy with $n\to \infty$, see \cite{Peschel2005} we get simply the result $S_{\infty}=\frac{\tilde{c}(\alpha)}{6}A(\alpha)$ which shows 
that in this case in contrast to the short-range case the single copy entanglement is not just half of the von Neumann entanglement entropy.

\begin{figure}[htp]
\begin{center}
\includegraphics[width=9cm,clip]{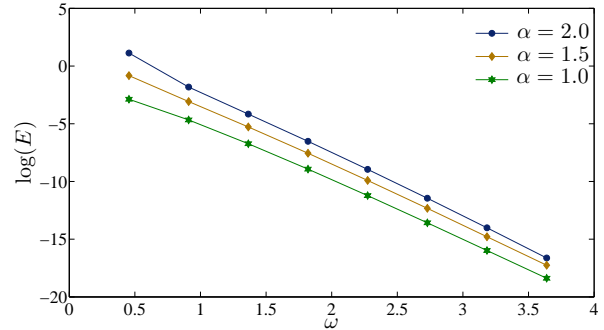}
\end{center}
\caption{(Color online) The eigenvalues $\log(E_i)$ versus $\omega_i$ for LRHO with the 
configuration $\mathfrak{C_4}$ and different $\alpha$s. The small eigenvalues 
(large $\omega_i$) are independent of $\alpha$ and also $\log(E_i)$ is linearly dependent to $\omega_i$.}
\label{Figure:7}
\end{figure}
In the next subsection, we will report the  results of LRHO in the case of 
a system which has a finite size and also  we will report the effect of boundary on the entanglement entropy.   

\subsection{Finite-size effects}

Until now to avoid any finite size effect, we concentrated on very large system
 size $L \rightarrow \infty$ and  small sub-system size $l$ (configuration $\mathfrak{C_1}$). 
As mentioned previously, the entanglement entropy $S$ and the R\'enyi entropy 
$S_n$, scale logarithmically with the size of the subregion $l$ ($l\ll L$). 
However, from the numerical computation of $\tilde{c}_n(\alpha)$, we argued that the LRHO 
is not conformally invariant except at $\alpha=2$.

We shall now present a computation of the entanglement entropy for 
systems with finite size. Conformal field theory 
(CFT) predicts\footnote{For more subtle FSE see \cite{Calabrese2010c,Xavier2012}.} following formulas for the R\'enyi entropy and the  von Neumann entropy of conformally invariant systems with
periodic BC's:
\begin{equation}\label{finiteEEVon}
S^{CFT}(L,l) = \frac{c}{3}\log \left[ \frac{L}{\pi} \sin \left(\frac{\pi l}{L} \right) \right]+c^\prime,
\end{equation} 
\begin{equation}\label{finiteEERen}
S_n^{CFT}(L,l) = \frac{c}{6}(1+\frac{1}{n})\log \left[ \frac{L}{\pi} \sin \left(\frac{\pi l}{L} \right) \right]+c^\prime_n ,
\end{equation} 
where $c$ is the central charge and  $c^\prime$ and $c^\prime_n$ are non-universal 
constants. Note that Eqs. (\ref{finiteEEVon}) and (\ref{finiteEERen}) are symmetric 
under $l \rightarrow L-l$, and they are maximal when $l = L/2$. For infinite system 
size $L\rightarrow \infty$ and also the finite one with the condition $l \ll L$ the 
entanglement entropy scales like Eq. (\ref{scalingEsmall_l}) \cite{Calabrese2004}. Notice that the
Eqs. (\ref{finiteEEVon}) and (\ref{finiteEERen}) are only true for conformally invariant 
systems and we expect different function in our system.

Here we will discuss the effect of boundary on the entanglement and R\'enyi 
entropies of the LRHO problem. We are interested to study the case, 
which we take a finite system with half of it as the sub-system. We considered 
a system with total size $2L$, and the sub-system size $L$ ($\mathfrak{C_4}$). The important subtility here
is the definition of the $K$ matrix. Since we have a finite system the fractional laplacian can not be easily defined by its 
Fourier transform (for more details see \cite{Zoia2007}). One way to define the fractional laplacian is based on non-local integrals 
in bounded domain \cite{Samko1993}. Although this approach is precise it is difficult to use it in discrete level for numerical evaluations. We will follow 
the simpler path the so called absorbing boundary condition considered in \cite{Zoia2007}.

The main difference between $K$ matrix in the finite system with boundary and the 
infinite one defined in \cite{Zoia2007} is that, the $K$ matrix for the system with boundary is defined by 
throwing away the elements of the infinite matrix which are in the outside of the system. 

Let us now consider the $\Lambda$ matrix and its eigenvalues $E$ for the 
configuration $\mathfrak{C_4}$. For the short range interaction problem 
($\alpha = 2.0$) the eigenvalues are described by $E = \sinh^{-2}(\pi\omega)$
 with $\omega = n\pi/\log(L)$ (see Eq. (\ref{energy callan})). Our calculations
 for other cases $\alpha <2$ show that the small eigenvalues are independent of
 $\alpha$ (see Fig. (\ref{Figure:7})). We found that $E = a(\alpha)/(\sinh^2(\pi\omega)+b(\alpha))$
(see also Eq. ($\ref{energy LRpSR}$)) is a good approximation for the eigenvalues of 
$\Lambda$ with $a(\alpha) = \frac{\alpha}{2}\sin^2 (\frac{\pi \alpha}{4})$ and 
$b(\alpha) =  0.32 \alpha -0.08 \alpha^2-0.16\alpha^3+0.06\alpha^4$ 
as the best numerical fit parameters to our data. The parameter $b(\alpha)$ for the 
configuration $\mathfrak{C_4}$ differs from the same quantity for the configuration $\mathfrak{C_1}$ except at $\alpha=2$.

\begin{figure}[htp]
\begin{center}
\includegraphics[width=9cm,clip]{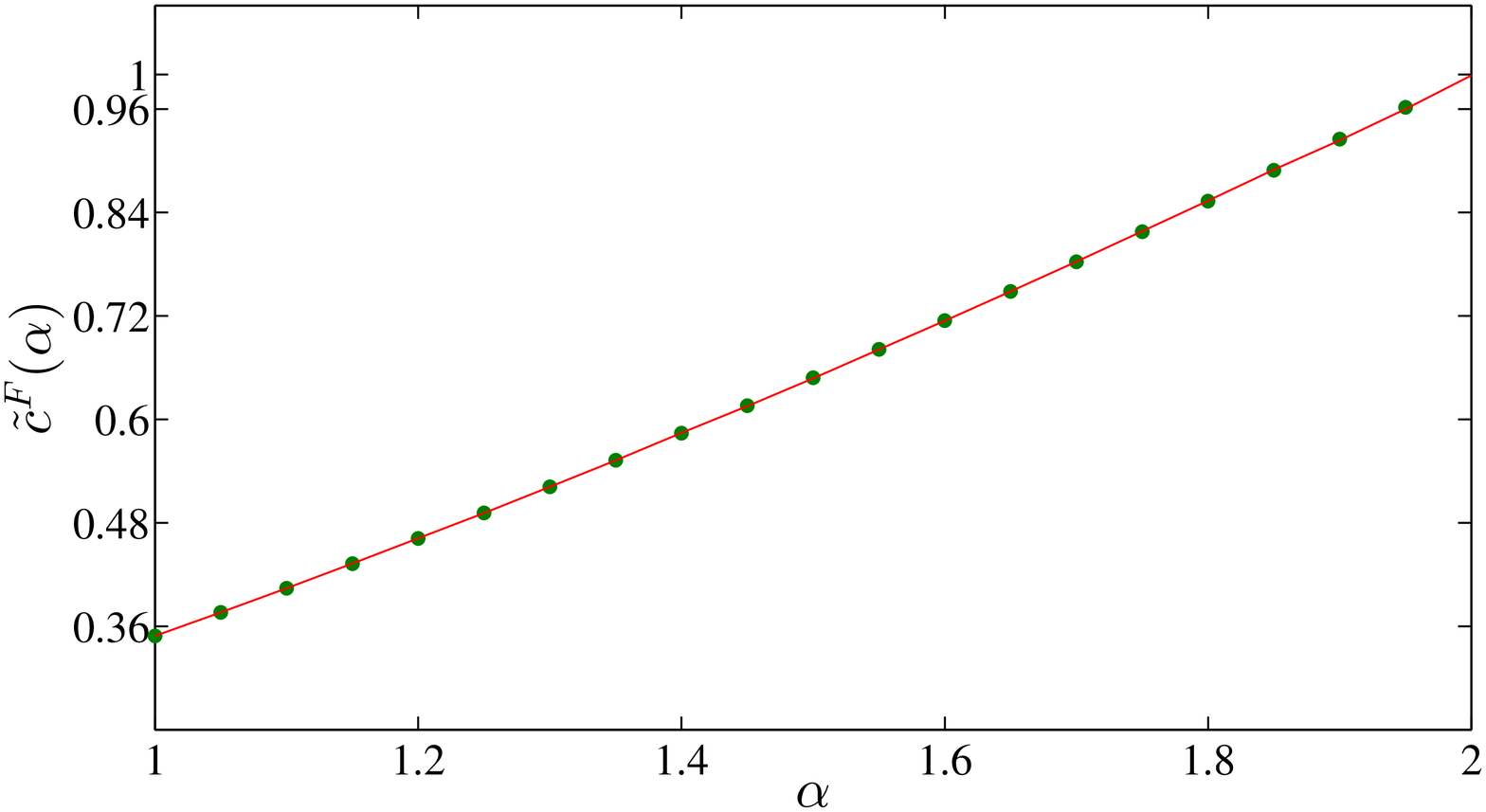}
\includegraphics[width=9cm,clip]{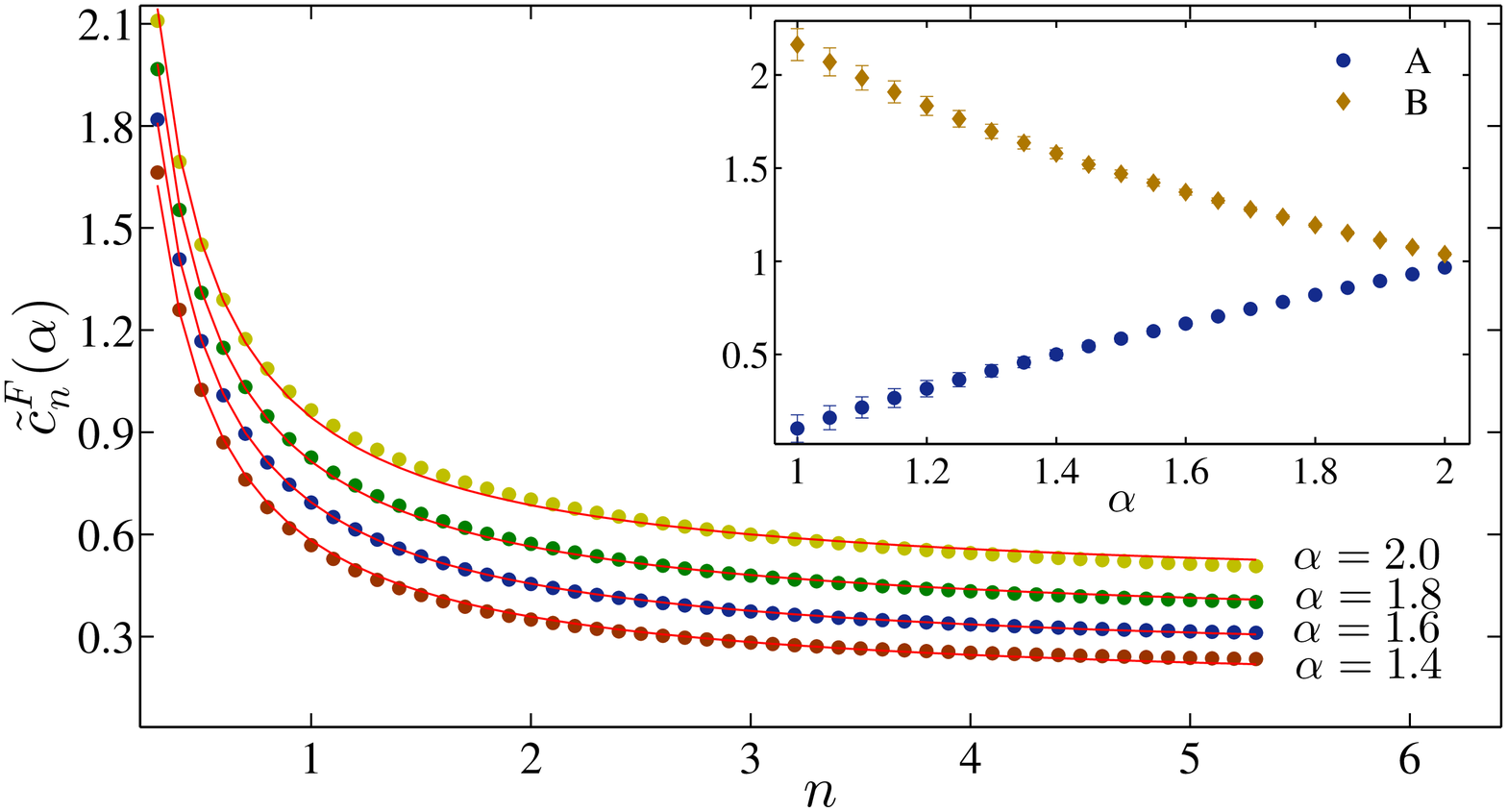}
\end{center}
\caption{(Color online) Top:  The scaling prefactor $\tilde{c}_4^F(\alpha)$ 
for discrete system with configuration $\mathfrak{C_4}$. The red line represents the same quantity 
coming from the continuum limit approximation. Bottom: $\tilde{c}^4_{4n}(\alpha)$ for the system with the 
configuration $\mathfrak{C_4}$, vs. $n$ for different $\alpha$'s (from top to bottom: $\alpha = 2.0, 1.8, 1.6, 1.4$). 
The red lines are the best fit with $\tilde{c}^{F}_{4n}(\alpha)=\frac{\tilde{c}^{F}_{4}(\alpha)}{2}(A^{F}(\alpha)+B^{F}(\alpha)/n)$. 
Inset: $A^F$ and $B^F$ coefficients vs. $\alpha$.}
\label{Figure:8}
\end{figure}

Numerical measurement shows that the entanglement entropy $S$ and the R\'enyi entropy $S_n$ follow
 $S\sim\frac{\tilde{c}_4^F(\alpha)}{6}\log L$ and 
$S_n\sim\frac{\tilde{c}^{F}_{4n}(\alpha)}{6}\log L$ respectively, where the  indices ~4 indicates the case that we study. In Fig. (\ref{Figure:8}) we report the numerically 
calculated values $\tilde{c}^F_4(\alpha)$ and $\tilde{c}^{F}_{4n}(\alpha)$ for several values of $\alpha$ and $n$. 

These prefactors are generally different from $\tilde{c}(\alpha)$ and $\tilde{c}_n(\alpha)$ except at the point $\alpha=2$.  
Finally we found that, $\tilde{c}^{F}_{4n}(\alpha)=\frac{\tilde{c}^{F}_4(\alpha)}{6}(A^{F}(\alpha)+B^{F}(\alpha)/n)$,
 is the best fit to $\tilde{c}^{F}_{4n}(\alpha)$ with respect to $n$ (notice that by definition $A^F(\alpha)+B^F(\alpha)=2$). 
The coefficients $A^F$ and also $B^F$ are functions of $\alpha$ (see Fig. (\ref{Figure:8})).

\begin{figure}[htp]
\begin{center}
\includegraphics[width=9cm,clip]{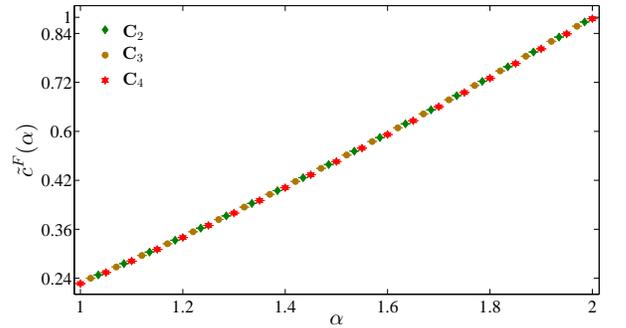}
\end{center}
\caption{(Color online) The scaling prefactor $\tilde{c}^F_i(\alpha)$ for discrete systems 
with configurations $\mathfrak{C_2}$ ,
 $\mathfrak{C_3}$ 
and $\mathfrak{C_4}$.}
\label{Figure:9}
\end{figure}

\begin{figure}[htp]
\begin{center}
\includegraphics[width=9cm,clip]{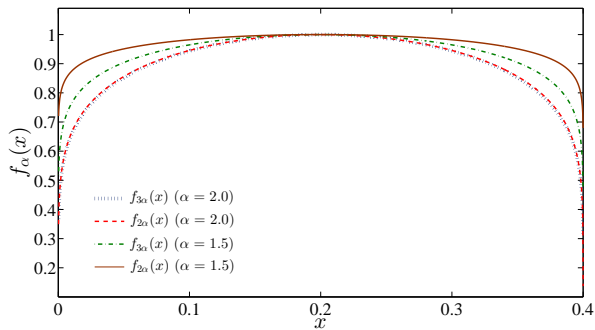}
\end{center}
\caption{(Color online) The function $f_{\alpha}(x)$ ($x=\frac{l}{L}$) for systems 
with configurations $\mathfrak{C_2}$ and $\mathfrak{C_3}$.}
\label{Figure:FSE f function}
\end{figure}
One can do the same calculations also for the configurations $\mathfrak{C_2}$ 
(in this case we considered $l=L/2$ and $S\sim\frac{\tilde{c}_2^F(\alpha)}{3}\log L$) and $\mathfrak{C_3}$
 (where we take $S\sim\frac{\tilde{c}_3^F(\alpha)}{6}\log L$). In Fig. (\ref{Figure:9}) 
we sketched $\tilde{c}^F(\alpha)$. It is clear that the results for different configurations 
$\mathfrak{C_2}$, $\mathfrak{C_3}$ and $\mathfrak{C_4}$ are similar. In other words
\begin{equation}
c^F(\alpha)=c^F_2(\alpha)=c^F_3(\alpha)=c^F_4(\alpha)
\end{equation}
In the next section 
we will discus  this similarity and we will show that these results are also the same as the massive systems. 
In case $\mathfrak{C_3}$ to have an idea about the function which controls the finite size effect we first realized that
one can fit the data to the following function

\begin{equation}
S =\frac{c^F_3(\alpha)}{6}\log (Lf_{3\alpha}(\frac{l}{L})),
\end{equation}
where $f_\alpha(x\to 0)\sim x$ and $f_\alpha(\frac{1}{2})\sim 1$. One can determine the function $f_\alpha$ by using 
the formula 
\begin{equation}
f_{3\alpha}(\frac{l}{L})=e^{\frac{6}{c^F_3(\alpha)}(S_\alpha(l)-S_\alpha(\frac{L}{2}))}.
\end{equation}
As one can see in Fig (\ref{Figure:FSE f function}) the function is smoothly $\alpha$ dependent. At the same Fig (\ref{Figure:FSE f function}) 
we also depicted the same function for the case
$\mathfrak{C_2}$ where we define $f_{2\alpha}(\frac{l}{L})=e^{\frac{3}{c^F_2(\alpha)}(S_\alpha(l)-S_\alpha(\frac{L}{2}))}$. It seems that except at the $\alpha=2$
the form of the functions are different in two different configurations.

\subsection{Massive LRHO}

As noted before, the entanglement entropy $S$ and the R\'enyi entropy $S_n$, in massless 
LRHO (for all configurations $\mathfrak{C_1}$, $\mathfrak{C_2}$, $\mathfrak{C_3}$ and $\mathfrak{C_4}$), 
increase logarithmically with the sub-system size. We also calculated the prefactors of the 
logarithms, $\tilde{c}(\alpha)$ and $\tilde{c}_n(\alpha)$ for the case $\mathfrak{C_1}$ and $\tilde{c}^F(\alpha)$ and
 $\tilde{c}^{F}_n(\alpha)$ for other cases, as a function of the long-range parameter $\alpha$ and $n$. 
Here we are interested in characterizing the massive long-range interacting harmonic oscillators. 

\begin{figure}[htp]
\begin{center}
\includegraphics[width=9cm,clip]{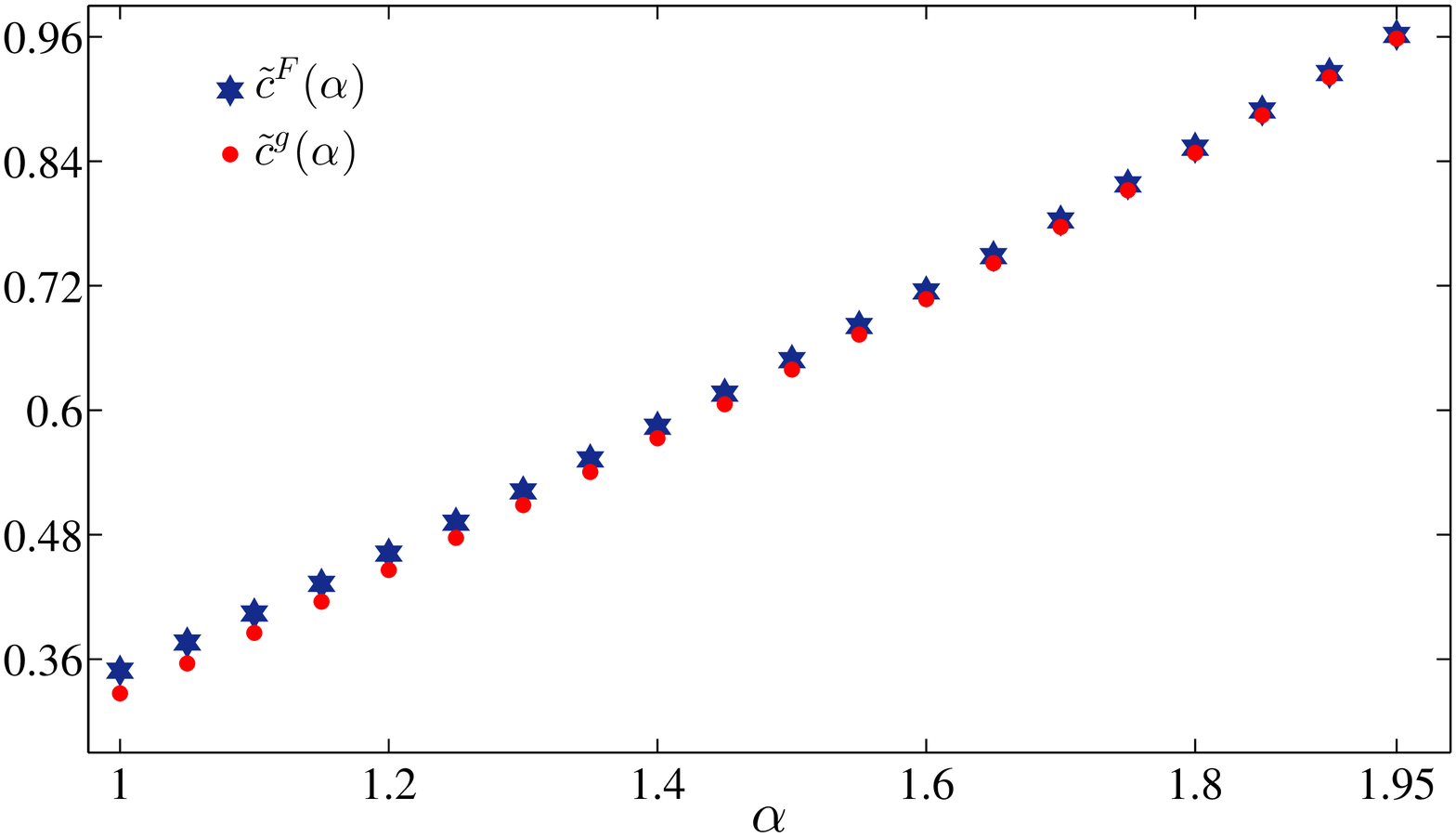}
\includegraphics[width=9cm,clip]{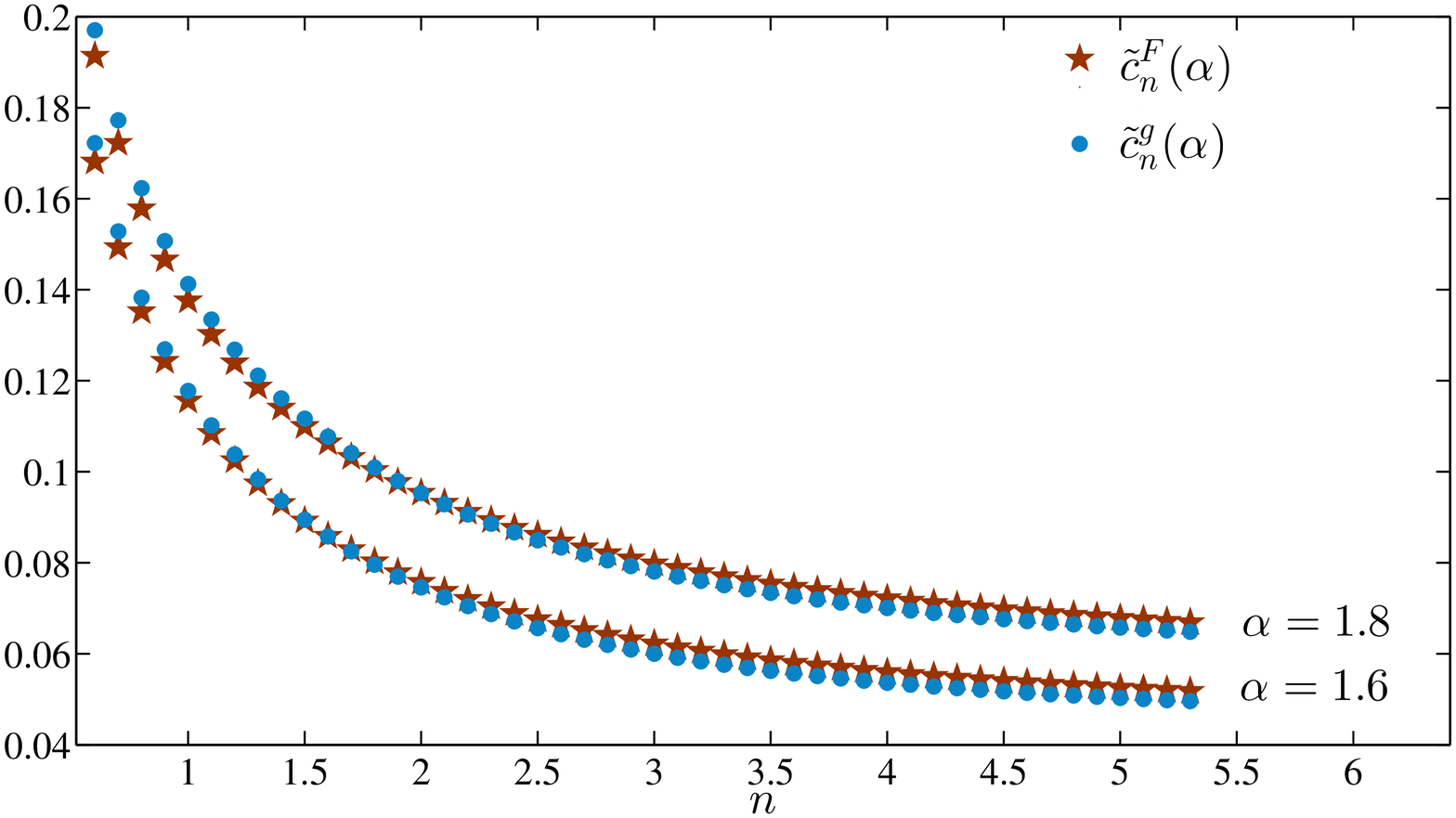}
\end{center}
\caption{(Color online)  Top: The prefactor $\tilde{c}^g(\alpha)$ compared with $\tilde{c}^F(\alpha)$ as function of $\alpha$. Bottom: The prefactor $\tilde{c}^{g}_n(\alpha)$ for massive LRHO with the configuration
 $\mathfrak{C_5}$ is the same as $\tilde{c}^{F}_n(\alpha)$ for massless one with the 
configurations $\mathfrak{C_2}$, $\mathfrak{C_3}$ and $\mathfrak{C_4}$.}
\label{Figure:11}
\end{figure}

First we consider a finite interval of length $l$ in a massive system 
(configuration $\mathfrak{C_5}$).
Following an argument given by Cardy-Calabrese \cite{Calabrese2004}, the entanglement entropy for such 
a system gets saturated by a mass scale and increases logarithmically $S =-\kappa\frac{c}{6}\log M$, 
where $c$ is the central charge of the system and it is equal to one for short-range harmonic oscillators 
and $M$ is the mass of the system. The prefactor $\kappa$ is the number of boundary points between 
subsystem $A$ and its complement with $\kappa=1$ for system with boundary and $\kappa=2$ for system
 with periodic boundary condition \cite{Calabrese2009}.  
 
We now consider the LRHO problem, Eq. (\ref{HOLR}) with mass $M>0$. As discussed before, we are again 
going to calculate the entropy $S$ numerically. 
The results, clearly show that $S$ saturates in the $l \rightarrow \infty$ and the entropy $S$ changes logarithmically with respect to the mass 
\begin{equation}
S =-\frac{\tilde{c}^g(\alpha)}{3}\log M~.
\end{equation}
Using such scaling form, we have obtained the prefactor $\tilde{c}^g(\alpha)$, as illustrated in
Fig. (\ref{Figure:11}), as a function of $\alpha$. Surprisingly we found that the prefactor 
is the same as the prefactor of the system with periodic boundary condition when we take half of the system (see  Fig.
 (\ref{Figure:11})). We also considered massive system with boundary and numerical results 
perfectly agree with $S =-\frac{\tilde{c}^g(\alpha)}{6}\log M$.

Next, we turn to speak more about the R\'enyi entropy $S_n$ for the massive case.  
Our analysis show that $S_n$ has also logarithmic scaling with the subsystem size and 
we measured the prefactor $\tilde{c}^{g}_n(\alpha)$. Interestingly we found that this exponent 
is the same as $\tilde{c}^{F}_n(\alpha)$. This is shown in Fig. (\ref{Figure:11}). To have an 
understanding of this equality we note that we generated the  $K$ matrix for the system with boundary ($M=0$) 
by throwing away those elements of the infinite system which are not inside the corresponding 
finite system. In this case the summation of the every row of the $K$ matrix is non-zero. 
This corresponds to an effective mass in the system and the system will be gapped. This effective 
mass is equivalent to the correlation length $\xi_t = \frac{1}{m^{\alpha/2}}$. Therefore, this 
argument hints that the results of massive LRHO should be similar to the massless one when the  system has boundary

\begin{figure}[htp]
\begin{center}
\includegraphics[width=9cm,clip]{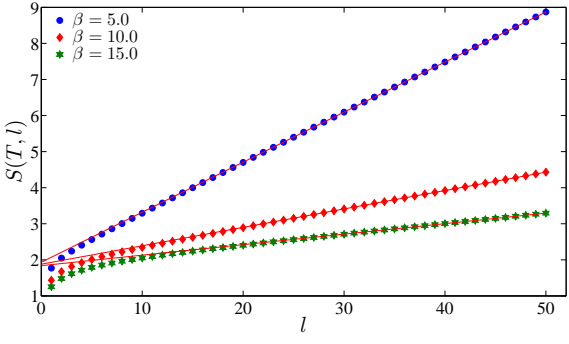}
\end{center}
\caption{(Color online) The von Neumann entropy for LRHO ($\alpha = 1.4$) with the system size $L=5000$, in the finite 
temperature $T = 1/\beta$.}
\label{Figure:13}
\end{figure}
\subsection{Finite temperature}\label{secEEfinitetemp}

In this section we present numerical results for the coupled harmonic oscillators 
with long range interaction in thermal states. Consider the Hamiltonian $\mathcal{H}$ 
at some temperature $T>0$. The Gibbs state corresponding to this temperature, associated 
with the canonical ensemble, is given by
\begin{equation}
\rho(\beta)=\exp(-\beta \mathcal{H})/\tr[\exp(-\beta \mathcal{H})],
\end{equation}
where $\beta=1/T$. Similar to the zero temperature case, one can obtain the covariance 
matrix $C(\beta)$ and also two point correlators $P(\beta)$ and $X(\beta)$, of 
the state $\rho(\beta)$ in the basis in which the Hamiltonian matrix is diagonal.
These matrices are given by \cite{Cramer2006}
\begin{equation}
P(\beta) = \frac{1}{2}K^{1/2}W(T), \hspace{1cm} X(\beta) = \frac{1}{2}K^{-1/2}W(T),
\end{equation} 
and
\begin{equation}\label{Comatthemp}
 C^2(\beta)=\frac{1}{4}(K^{-1/2}W(T))\oplus(K^{1/2}W(T)),
\end{equation} 
where $W(T) :=\mathbb{I}+2(\exp(K^{1/2}/T)-\mathbb{I})^{-1}$. 
It is worth mentioning that the entropy of the
subsystem with length $l$ at temperature $T$ for CFT is given by the formula \cite{Calabrese2004}:
\begin{equation}\label{finite temp EE}
S =\frac{c}{3}\log \left( \frac{\beta}{\pi}\sinh\frac{\pi l }{\beta} \right)+c^\prime_1 ~.
\end{equation}
\begin{figure}[htp]
\begin{center}
\includegraphics[width=9cm,clip]{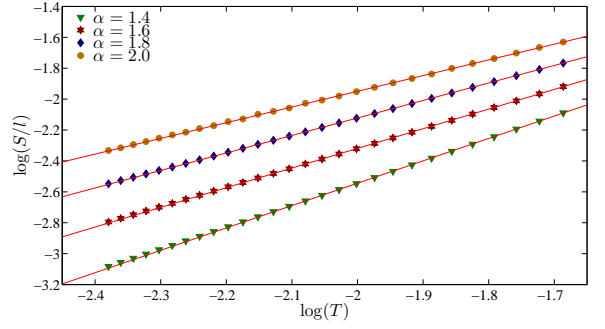}
\includegraphics[width=9cm,clip]{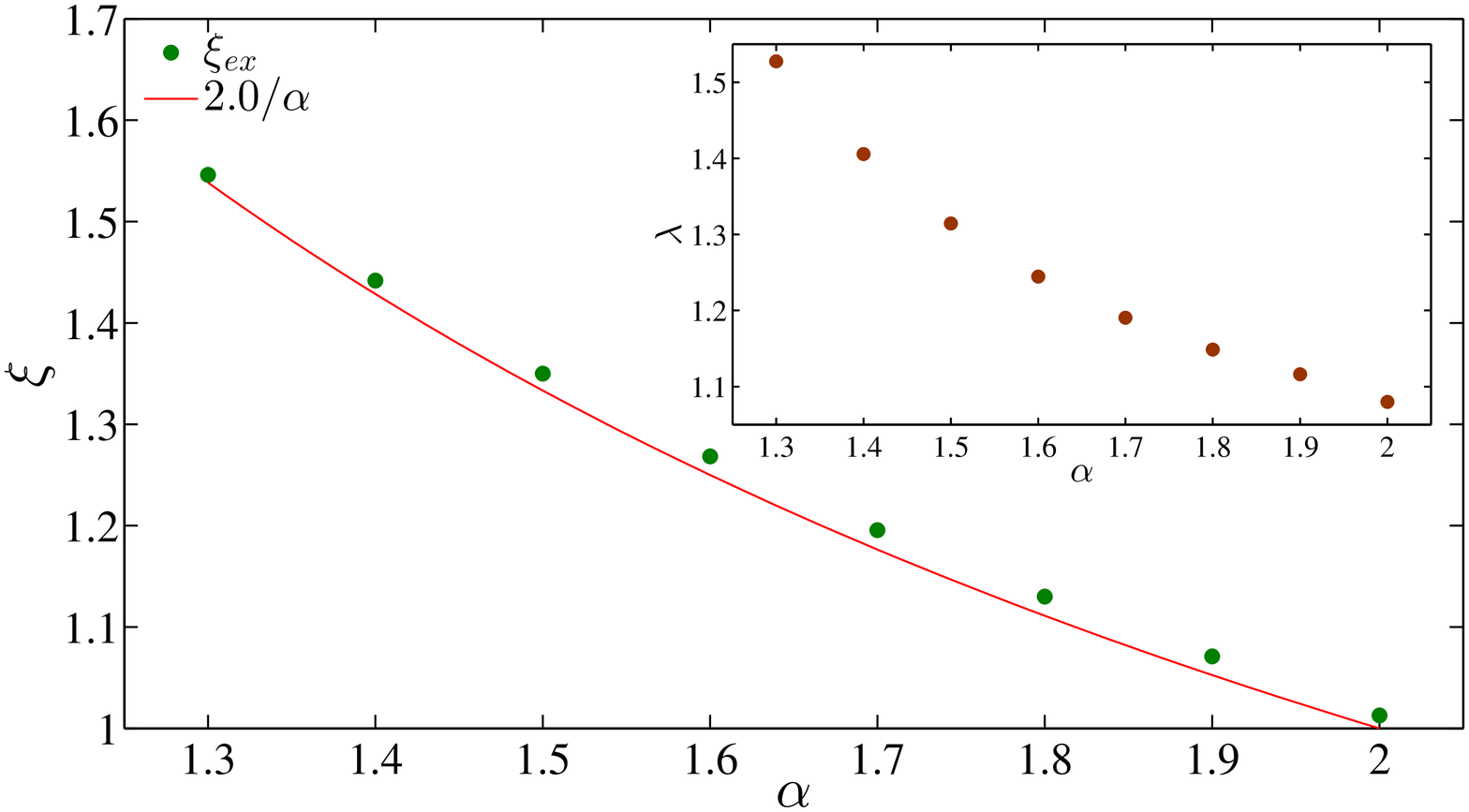}
\includegraphics[width=9cm,clip]{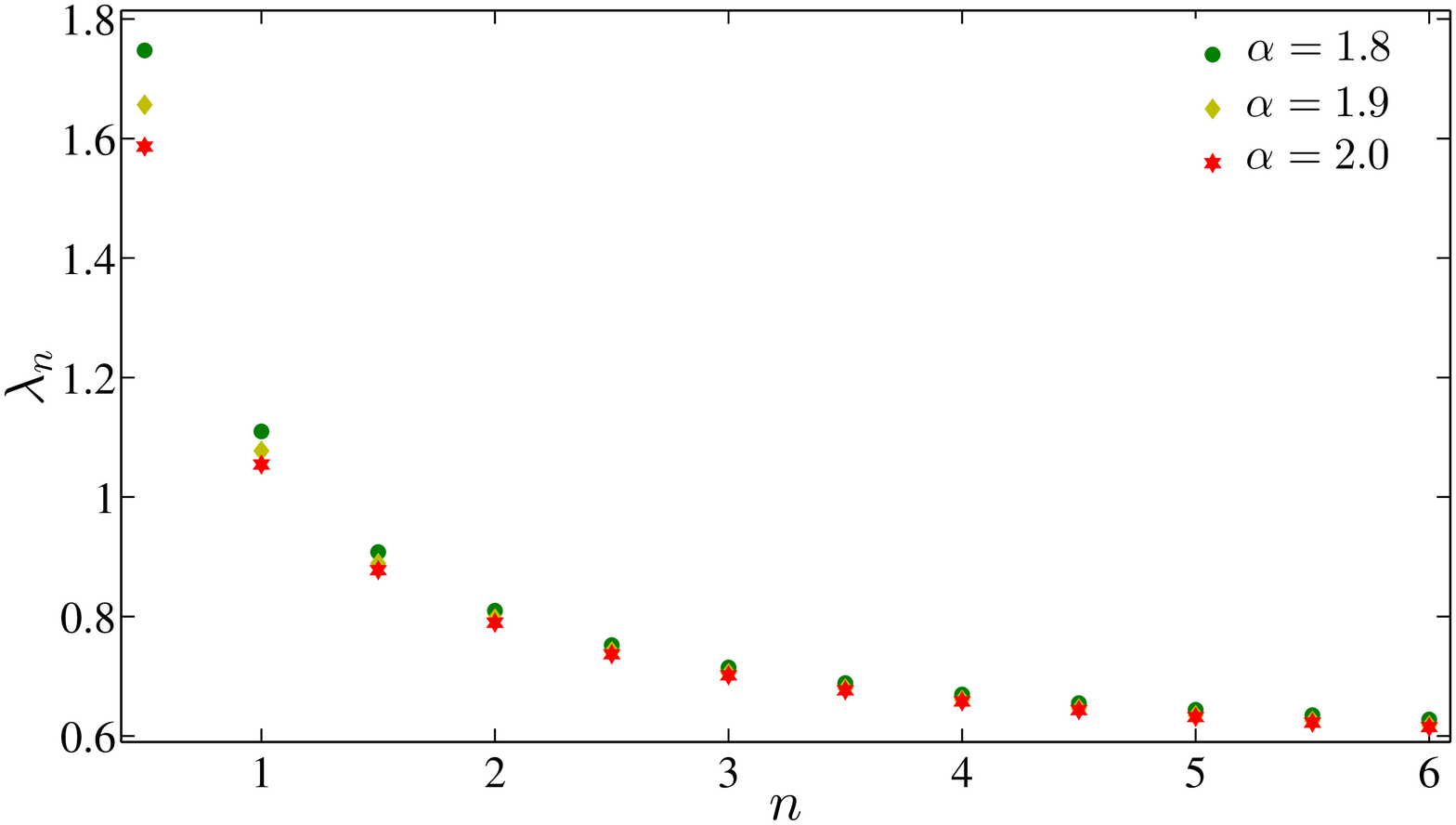}
\end{center}
\caption{(Color online) Top: The entanglement entropy for LRHO with finite temperature 
shows the scaling behavior $S(T,l)/l \sim \lambda T^\xi$, in the high temperature limit.
 Middle: The scaling exponent $\xi$ as function of $\alpha$. The solid red line represents 
$\xi = 2/\alpha$. The parameter $\lambda$ is shown in the inset. Bottom: The parameter 
$\lambda_n$ for the R\'enyi entropy of LRHO with finite temperature  in the high temperature limit.}
\label{Figure:14}
\end{figure}
As expected, in the zero and  high temperature limits, the von Neumann entropy reduces to 
$S=\frac{c}{3}\log {l}+c^\prime_1$ and $S = \frac{\pi c}{3\beta}l+c^\prime_1$, respectively.
 In the high-temperature limit, the von Neumann entropy has an extensive form and it reduces
 to the standard CFT and agrees with the Gibbs entropy of an isolated system of length $l$ \cite{Calabrese2004}. 

\begin{figure} [htp]
\centering
\includegraphics[width=9cm,clip]{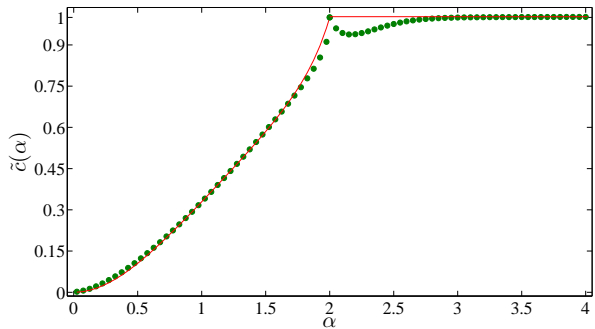}
\caption{(Color online) Green (Dark gray) circles represent $\tilde{c}(\alpha)$ 
for the system of harmonic oscillator with long range plus short range interaction
 with the configuration $\mathfrak{C_1}$. The prefactor $\tilde{c}(\alpha)$ is 
measured using the scaling relation $S$ with $\log l$ in the region $0<l<L/100$ 
for the system size $L=6000$. The red line represents the same quantity for the pure LRHO. }
\label{Figure:long plus short1}
\end{figure}

As in previous cases, our aim is to study the properties of the von Neumann entropy of 
system of harmonic oscillators with long range interaction at finite temperature. For simplicity 
we focus on high temperature limit. In order to measure the von Neumann entropy, we needed to 
restrict the system size to the finite values with total size $L$, and subsystem size $1\ll l \ll L$, to avoid finite size problem. In order to calculate the von Neumann in this state, we need to 
consider the covariance matrix Eq. (\ref{Comatthemp}) associated with the reduced state of an interval 
with length $l$. Thus, we calculated $C(\beta)$ at some particular values $T=1/\beta$ and then performed 
the diagonalization of the covariance matrix to find $S(T,l)$.

As shown in Fig. (\ref{Figure:13}), $S(T,l)$ for various values of $T$ and $\alpha$, in the high-temperature limit is 
a linear function of $l$, so in this case one has
\begin{eqnarray}
S \sim \lambda l T^\xi 
\end{eqnarray}
where $\lambda$ and $\xi$ are functions of $\alpha$. The log-log plot $S/l$ with respect to $T$ is shown in 
the Fig. (\ref{Figure:14}).      
The scaling parameter $\xi$ and the prefactor $\lambda$ are shown in Fig. (\ref{Figure:14}). The scaling 
exponent $\xi$ and the quantity $2/\alpha$ are the same, and one can nicely interpolates $\xi=2/\alpha$.
 This is not surprising because it is well known that in the long-range systems the dynamical exponent 
is $z =\frac{\alpha}{2}$ and this exponent controls the relative scaling of time and space leading to the
 invariant form $lT^{1/z}$ \cite{Swingle2011}.
In  general the thermal entropy for the theories with the dynamical exponent $z\neq 1$ scale as $lT^{1/z}$ 
which it follows from the requirement of dimensionlessness and extensivity \cite{Swingle2011}. Returning to our 
LRHO problem we can conclude that the entropy in high temperature limit should follow the simple 
form $S\propto lT^{2/\alpha}$.

It is interesting to note that the R\'enyi entropy $S_n$ for finite temperature LRHO, in the high-temperature limit is 
\begin{eqnarray}
S_n \sim \lambda_n l T^\xi. 
\end{eqnarray}
In Fig. (\ref{Figure:14}) we show the prefactor $\lambda_n$ as a function of $n$ for several $\alpha$'s. 
It is worth mentioning that all the curves have similar behavior at large $n$ ($\lambda_{\infty}=\pi/6$).

\subsection{Universality}

In the previous sections we studied a particular case of long-range harmonic oscillator which leads to a well-defined 
continuum limit field theory. This is a hint to believe that probably the results that we found are robust and valid 
for larger set of harmonic oscillators. In this section we would like to address this question by first studying
long-range harmonic oscillator in the presence of short-range harmonic oscillator and then by investigating larger set of 
interactions which can be decomposed to our studied systems. 
\begin{figure} [htp]
\centering

\includegraphics[width=9cm,clip]{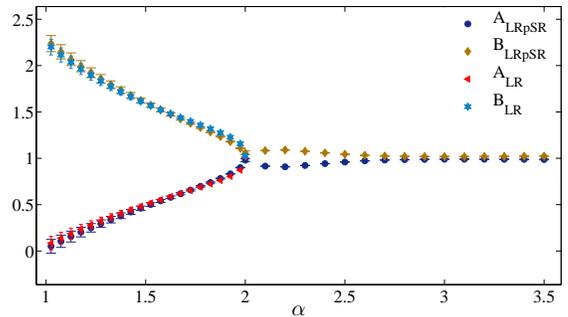}

\caption{(Color online) $A(\alpha)$ and $B(\alpha)$ coefficients versus $\alpha$ for system of harmonic oscillator with long range plus short range interaction.}
\label{Figure:long plus short2}
\end{figure}

\subsubsection{Long-range HO in the presence of short-range HO}

So far, we have only considered the harmonic oscillator systems with the long-range interaction. 
In the last section, we studied LRHO by means of eigenvalue problem and we computed the
 eigenvalue $E$ and the eigenfunction $\psi$ numerically. 
In a similar way we will try to do the same calculation for harmonic oscillator systems
 with long-range plus short-range interaction. Then, we will study the logarithmic 
scaling of the entanglement entropy $S$ and also R\'enyi entropy $S_n$ for this model. 
Finally, we are going to analyze the scaling coefficient $\tilde{c}(\alpha)$ and $\tilde{c}_n(\alpha)$ as  functions of $\alpha$.

Consider the hamiltonian Eq. (\ref{harmonicOsc}), with long-range plus short-range interaction:
\begin{equation}\label{SRpLRHO}
K = K_{LR}+K_{SR},
\end{equation}
where $K_{LR}$ is again defined as Eq. (\ref{HOLR}), and $K_{SR}$ is just a simple laplacian. 
We have only considered the massless system with $M=0$ but one can also generalize them to $m\neq 0$.

We have carried out simulations for $0<\alpha<4$. For each value of $\alpha$, we 
have determined the matrix $K$ for the large enough system size with $L=6000$ with
 the sub-system size less than $L/100$. The entanglement entropy grows logarithmically with the sub-system 
size as $S=\frac{\tilde{c}(\alpha)}{3}\log(l)$, where $\tilde{c}(\alpha)$ is a function of $\alpha$. 
We have depicted $\tilde{c}(\alpha)$ versus $\alpha$ in Fig. (\ref{Figure:long plus short1}), where the solid 
line comes from LRHO case. It is also interesting to note the similarity of 
$\tilde{c}(\alpha)$ in the range $\alpha<2$ with the results of harmonic oscillator with pure 
long-range interaction and also $\alpha\geq 2$ with the result of HO with pure short-range 
interaction (see Fig. (\ref{Figure:long plus short1})). The entanglement entropy of harmonic oscillator 
system with long range plus shot range interaction with the exponent $\alpha<2$ ($\alpha \geq 2$)
 is the same as harmonic oscillator system with pure long-range (short-range) interaction. This might look not surprising 
because we know that from the renormalization group point of view the short-range interaction is irrelevant as far as $\alpha<2$. 
In our numerical calculation the reason of discrepancy in the region $2<\alpha\leq 2.5$ is unclear to us. 

\begin{figure}[htp]
\begin{center}
\includegraphics[width=9cm,clip]{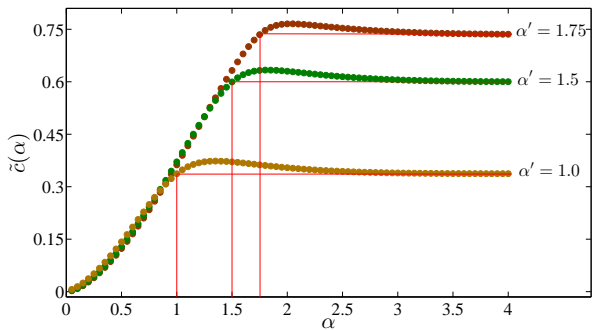}
\end{center}
\caption{(Color online) The prefactor $\tilde{c}(\alpha)$ for a system of harmonic oscillator with long range interaction with exponent $\alpha$ plus another long range interaction with the exponent $\alpha^\prime$. It seems that $\tilde{c} \sim \min\lbrace \tilde{c}(\alpha),\tilde{c}(\alpha^\prime)\rbrace$. The prefactor $\tilde{c}(\alpha)$ is measured using the scaling relation $S$ with $\log l$ in the region $0<l<L/100$ for the system size $L=6000$. }
\label{Figure:long plus long range}
\end{figure}

We also calculated the R\'enyi entropy $S_n$ for coupled harmonic oscillators with long-range 
plus short range couplings. To get $S_n$ numerically, we used  Eq. \ref{Renyi}. We found that 
for $l\ll L$ the R\'enyi entropy also logarithmically increases with the system size as 
$S_n=\frac{\tilde{c}_n(\alpha)}{3}\log(l)$, where the prefactor $\tilde{c}_n(\alpha)$ is a function of $n$ and also $\alpha$. 
The best fit is $\tilde{c}_n(\alpha)=\frac{\tilde{c}(\alpha)}{2}(A(\alpha)+B(\alpha)/n)$. The resulting values of $A(\alpha)$ and $B(\alpha)$ as a
function of $\alpha$
are represented in Fig. (\ref{Figure:long plus short2}). We remark that, for $\alpha<2$ the data are in excellent 
agreement with the LRHO \cite{Nezhadhaghighi2012}, whereas for $\alpha\geq2$ they agree with the short range prediction. 
On the other hand, the system is conformally invariant for $\alpha\geq 2$ where we have $A=B=1$.

We now consider the entanglement entropy of a system of long-range harmonic oscillator 
with $K = K_{LR}^{\alpha}+K_{LR}^{\alpha^\prime}$, where $K_{LR}^{\alpha}$  is defined 
as in Eq. (\ref{HOLR}). The entanglement entropy grows logarithmically with
 the sub-system size and the prefactor is equal to $\tilde{c} \sim \min\lbrace \tilde{c}(\alpha),\tilde{c}(\alpha^\prime)\rbrace$.
 The results of the prefactor $\tilde{c}$ is depicted in the Fig. (\ref{Figure:long plus long range}). For $\alpha \geq \alpha^\prime$ 
we expect $\tilde{c}\sim \tilde{c}(\alpha^\prime)$ but when $\alpha \sim \alpha^\prime $ we observe a large discrepancy in the numerical results.

\subsubsection{Generalization to singular Toeplitz matrices}\label{Generalized entropy}
In this subsection we would like to address how one can relate the entanglement entropy of more general harmonic oscillators
to the entanglement entropy of the studied long-range harmonic oscillators. Although our conclusion is based on just numerical evaluations
 we will show in the section dedicated to the mutual information that in some particular cases one can derive the results 
analytically. We define the hamiltonian of the harmonic oscillator  with the following $K$ matrix:
\begin{figure}[htp]
\begin{center}
\includegraphics[width=9cm,clip]{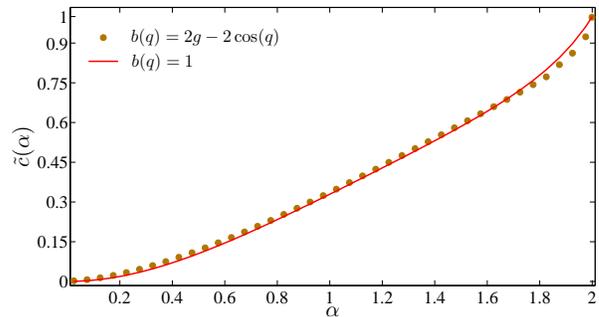}
\end{center}
\caption{(Color online) The prefactor of logarithm for the entanglement entropy for non-trivial $b(q)$ 
function. For $g>1$ the prefactor is independent of the $b(q)$ function. }
\label{Figure:universalityofbP}
\end{figure}
\begin{eqnarray}\label{Toeplitz1}
K_{l,m}= - \int _{0}^{2\pi} \frac{dq}{2\pi}e^{iq(l-m)} {b(q)\prod_{r=1}^{R}u(\alpha_r,q-q_r)},
\end{eqnarray}
where $b(q):S^1\to \mathcal{C}$ is a smooth non-vanishing function with zero winding number and 
\begin{eqnarray}\label{u function1}
u(\alpha,q)=(2-2\cos q)^{\frac{\alpha}{2}}.
\end{eqnarray}

The above Toeplitz matrices are usually called singular Toeplitz matrices. For our purpose we need to also consider some
particular restrictions on $q_r$ to have real interactions between harmonic oscillators. From now on we will consider just 
those $q_r$'s that $e^{i q_r}$'s are either real or the complex conjugate of each other for every $\alpha_r$. Harmonic oscillators 
with the above interactions are critical and one can simply show that $\xi^{-1}=0$. The above interactions are natural generalizations 
of the ones that we considered in previous sections. One way to see this is by considering $2m-2\cos q$ instead of $2-\cos q$ in the 
equation (\ref{HOLR}). For $m=\cos q_r$ one can simply show that $|2m-2\cos q|=(2-2\cos(q+q_r))^{\frac{1}{2}}(2-2\cos(q-q_r))^{\frac{1}{2}}$
which is in the form of (\ref{Toeplitz1}). It is worth mentioning that for $m>1$ the system is gapped and otherwise it is gapless. 

\begin{figure}[htp]
\begin{center}
\includegraphics[width=9cm,clip]{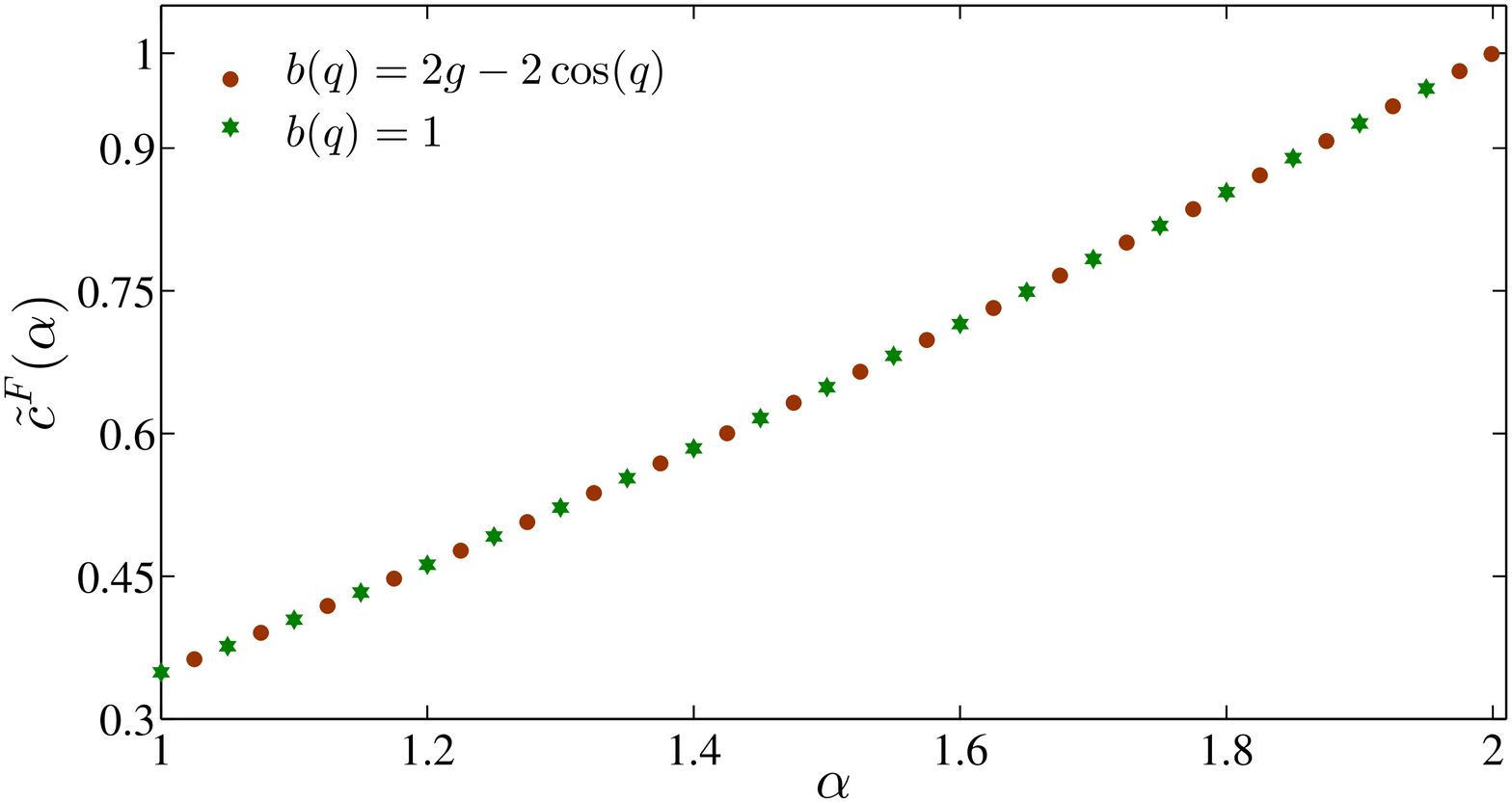}
\end{center}
\caption{(Color online) The prefactor of logarithm in the presence of boundary for non-trivial $b(q)$ function. For $g>2$ the prefactor is independent of the $b(q)$ function. }
\label{Figure:universalityofbF}
\end{figure}

Using the techniques of 
the previous sections one can calculate easily the entanglement entropy of these harmonic oscillators. The entanglement entropy
changes logarithmically with the subsystem size and the prefactor of the logarithm is a function which is independent of $b(q)$
and $q_r$ but it is strongly dependent on the $\alpha_r$'s. To show that the results are $b(q)$ independent we took 
$b(q)=2g-2\cos(q)$ with $g>1$ for several $g$ for $R=1$. The results are shown in the Fig.(\ref{Figure:universalityofbP}) and
 Fig(\ref{Figure:universalityofbF}) where one can see that the
 prefactor of the logarithm is the same in all the different cases.  We conjecture that the prefactor of the logarithm
is independent of the form of the function $b(q)$. Next we calculated the prefactor of the logarithm for different values of
$\alpha_r$ and $q_r$. The results are shown in the table ~\ref{tab1}. 

\begin{table}[htp]
\begin{center}
\begin{tabular}{l*{15}{c}r}    
\hline\hline
$q_1$ & $q_2$ & $q_3$ & $q_4$ & $q_5$& $q_6$& $\alpha_1$ & $\alpha_2$ & $\alpha_3$ & $\alpha_4$ & $\alpha_5$& $\alpha_6$&  ${\tilde{c}}/{2}$\\ \hline 
$\frac{\pi}{3}$& $\frac{-\pi}{3}$ & 0 & 0 & 0  & 0 & 1 & 1 & 0  & 0  & 0  & 0  &  0.33(0.01) \\ 
\hline
$\frac{\pi}{6}$& $\frac{-\pi}{6}$ & 0 & 0 & 0  & 0 & 1 & 1 & 0  & 0  & 0  & 0  &  0.33(0.01) \\ 
\hline
0 & 0 & $\frac{\pi}{3}$ & $\frac{-\pi}{3}$ & 0 & 0  & 0 & 0 & 1.5 & 1.5  & 0 & 0  &  0.60(0.01) \\  
\hline
0& 0  & 0 & 0 & $\frac{\pi}{3}$  & $\frac{-\pi}{3}$ & 0 & 0 & 0  & 0  & 2  & 2  &  0.99(0.01) \\  
\hline  
$\frac{\pi}{3}$ & $\frac{-\pi}{3}$ & $\frac{\pi}{6}$ & $\frac{-\pi}{6}$ & 0  & 0 & 1 & 1 & 1.5  & 1.5  & 0  & 0  &  0.92(0.01) \\ 
\hline
$\frac{\pi}{3}$ & $\frac{-\pi}{3}$ & $\frac{\pi}{6}$ & $\frac{-\pi}{6}$ & 0  & 0 & 1 & 1 & 1.5  & 1.5  & 2  & 2  &  1.91(0.01) \\  
\hline     
$\frac{\pi}{3}$ & $\frac{-\pi}{3}$ & $\frac{\pi}{6}$ & $\frac{-\pi}{6}$ & $\frac{\pi}{4}$  & $\frac{-\pi}{4}$ & 1 & 1 & 1.5  & 1.5  & 2  & 2  &  1.93(0.01) \\ 
    \hline \hline
    \end{tabular} 
\end{center}
\caption{\label{tab1} Numerical values of the prefactor ${\tilde{c}}$ for different values of
$\alpha_r$ and $q_r$.}
\end{table}

It is easy to see that firstly the results are independent of $q_r$'s and secondly one 
can get the results of the last three rows by just summing the results of the first four
 rows.
Based on the results of the table one can  conjecture that for the interaction (\ref{Toeplitz1}) 
the following result is valid for the prefactor of the logarithm:
\begin{eqnarray}\label{c decomposition}
\begin{split}
&\tilde{c}(\alpha_1,\alpha_1,\alpha_3,\alpha_3,\dots,\alpha_{R-1},\alpha_{R-1})=\\
&\tilde{c}(\alpha_1,\alpha_1,\dots,0)+\tilde{c}(0,0,\alpha_2,\alpha_2,\dots,0)+\dots\\
&+\tilde{c}(0,0,\dots,\alpha_{R-1},\alpha_{R-1}).
\end{split}
\end{eqnarray}

In other words one can get the prefactor of the logarithm in the model (\ref{Toeplitz1}) by just having the same 
quantities for the long-range harmonic oscillator that we have discussed in the previous sections
\subsection{Two dimensions: area law and logarithmic term for polygonal region}

It was shown in \cite{Srednicki1993} that the area law is valid for short-range harmonic oscillator if we consider a sphere like region in higher dimensions.
The coefficient of the area term is a non-universal number. For example, if we take an square like subregion then the coefficient of the area term
 will be dependent to the orientation of the
polygon with respect to the symmetry axes of the lattice. Later it was shown in \cite{Casini2006} that if we consider a region with sharp corners then
in the entanglement entropy there will be also some extra logarithmic terms with universal coefficients. In other words one can write the entanglement entropy of a polygon as
\begin{figure} [htp]
\centering
\includegraphics[width=9cm,clip]{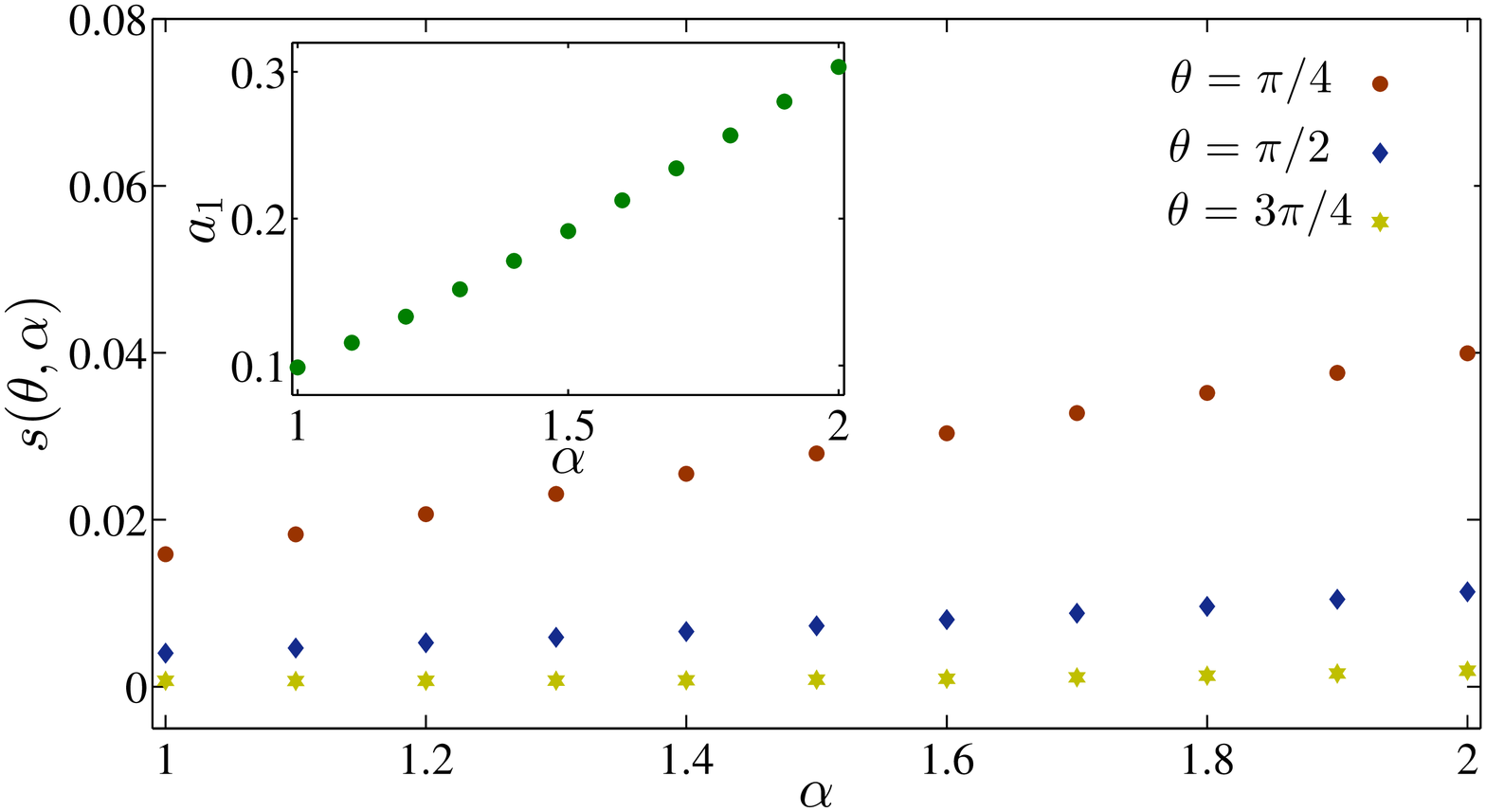}
\caption{(Color online) $s(\theta,\alpha)$ as a function of $\alpha$ for subsystems with different vertex angles $\theta$. \textit{Inset}:
the nonuniversal coefficient of the area term $a_1$ with respect to different values of $\alpha$. }
\label{Figure:EE_2d_area_and_log_coeff}
\end{figure}
\begin{eqnarray}\label{EE polygon}
S(\theta)=a_0+a_1 L+a_{-1}L^{-1}+a_{-2}L^{-2}-s(\theta,\alpha)\log L,
\end{eqnarray}

where $L$ is the size of the system and $\theta$ is the vertex angle of the polygon. Following the same procedure as previous sections we first found 
the entanglement entropy of the square like regions for different values of $\alpha$'s and confirmed that the leading term is the area law. The coefficient $a_1$
was an increasing function of $\alpha$ (See Fig.(\ref{Figure:EE_2d_area_and_log_coeff})). Using the equation (\ref{EE polygon}) then we found $s(\theta,\alpha)$ for different values of $\theta$ such as: 
$\theta=\frac{\pi}{4},\frac{\pi}{2},\frac{3\pi}{4}$ and different values of $\alpha$. The results are depicted in 
Fig.(\ref{Figure:EE_2d_area_and_log_coeff}) We also showed that 
the coefficients $s(\theta,\alpha)$ are independent of the orientation of the subregion with respect to the symmetry axes of the lattice. One can summarize
this section as follows: the entanglement entropy of a polygonal region for long-range harmonic oscillators follows the same formula
as the short-range one but with different coefficients. We also confirmed that the same kind of behavior is also valid for R\'enyi entropy.

\section{Mutual information}

In the previous sections we studied the von Neumann entropy $S$ and the 
R\'enyi entropy $S_n$ for long range harmonic oscillators with different 
configurations of systems and subsystems. It is also of considerable interest 
to quantify the Shannon's classical mutual information \cite{Wolf2008} for  system of 
harmonic oscillators with short range and long range interactions. The Shannon information for spin systems 
were first studied in \cite{Stephan2009} and much more investigated in \cite{Othersmutual1,Wilm2011,Um2012,Alcaraz2013} for different quantum systems. Here we focus to the definitions
given in \cite{Cramer2006,Unanyan2005,Alcaraz2013}.

\begin{figure}[htp]
\begin{center}
\includegraphics[width=9cm,clip]{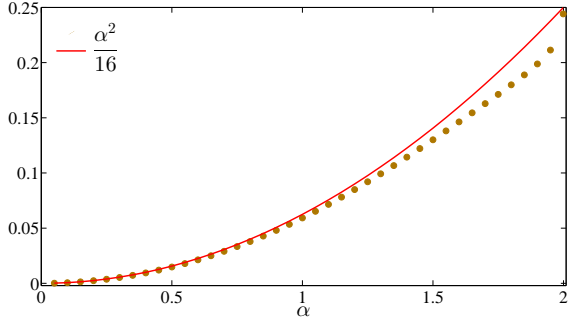}
\end{center}
\caption{(Color online) The prefactor of the logarithmic scaling of the mutual information $I_1$ for LRHO with configuration $\mathfrak{C_1}$.}
\label{Figure:mutual pbc small size 1}
\end{figure}

Consider a chain of $L$ harmonic oscillators described by canonical 
variables ($\phi_i, \pi_i$), $i=1, 2, \dots,L$, and the system is divided
in to two parts $A$ and $B$ with $l$ and $L-l$ oscillators, respectively. 
The oscillators are coupled by a quadratic hamiltonian Eq. (\ref{harmonicOsc}). 
Let us now consider $\Phi {_A} = (\phi_1, \phi_2, \dots, \phi_{l})$ and $\Phi_B = (\phi_{l+1}, \phi_{l+2}, \dots, \phi_L)$ 
the position vectors of the subsystems $A$ and $B$ and $\Pi_{A,B}$ the respective momentum vectors. 
The classical mutual information can be defined as:
\begin{eqnarray}
I (A,b)= S_A +S_B -S_{(A+B)}~,
\end{eqnarray}
where $S$ is the Shannon's classical entropy. There are in fact two different definitions to evaluate Shannon's mutual information. The difference 
comes from the source of probabilities that we use to define the entropy. In the first case we use the ground state of the quantum system as the source of
the probabilities for appearing different configurations and in the second case it will be just the Gibbs distribution. The first definition which has recently found
many interesting applications in the study of spin chains \cite{Stephan2009,Wilm2011,Um2012,Alcaraz2013} can be defined in the context of
harmonic oscillators as follows:
The Shannon's classical mutual information $I(A:B)$ between two regions $A$ and $B$ is  

\begin{eqnarray}\label{mutual1}
I_1(A:B) = \int d^N\Phi p(\Phi_A,\Phi_B)\ln \frac{p(\Phi_A,\Phi_B)}{p(\Phi_A)p(\Phi_B)} 
\end{eqnarray}
where $p(\Phi_A,\Phi_B)=\vert \Psi_0\vert ^2$ is the total and $p(\Phi_{A})=\int \left[\prod_{m\in ({B})}d\phi_m\right] 
\vert \Psi_0\vert^2$ and $p(\Phi_{B})=\int \left[\prod_{m\in ({A})}d\phi_m\right] 
\vert \Psi_0\vert^2$ are the reduced probability densities in position space ($\Psi_0$ is the ground state 
wave function i.e. Eq. (\ref{GroundSwave})) \cite{Unanyan2005}.

 The mutual information $I(A:B)$ or $I(A:B)$ quantifies how correlated two parts 
are when the system is in the ground state and for harmonic oscillators has the following simple form:
 
\begin{figure}[htp]
\begin{center}
\includegraphics[width=9cm,clip]{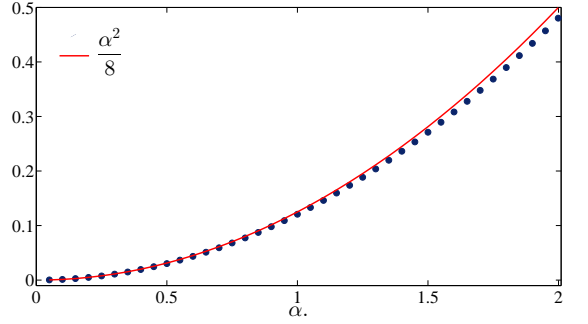}
\end{center}
\caption{(Color online) The prefactor of the logarithmic scaling of the mutual information $I_2$ for LRHO 
with configuration $\mathfrak{C_1}$.}
\label{Figure:mutual pbc small size 2}
\end{figure}
\begin{eqnarray}\label{mutual2}
\begin{split}
I_1(A:B) &= \frac{1}{2}\ln \frac{(\det 2X_A)(\det 2X_B)}{\det K^{-1/2}} \\ 
&= \frac{1}{2}\ln \frac{(\det 2P_A)(\det 2P_B)}{\det K^{1/2}}  \\ 
& = \frac{1}{2}\ln (\det 4 X_A P_A) = \sum_{i=1}^{l} \ln 2\nu_i ~,
\end{split}
\end{eqnarray} 
where $X_A$ and $P_A$ are $l$ dimensional matrices describing correlations within a 
compact block of $l$ oscillators (subsystem $A$) and $\nu_i$ is the eigenvalue of 
the matrix $C = \sqrt{X_A P_A}$ and $X_B$ and $P_B$ are $(L-l)\times(L-l)$ matrices
 describing correlations within subsystem $B$, and the matrices $X_{AB}$ and $P_{AB}$ 
describe the correlations between them (see Eq. (\ref{X_A P_A})) \cite{Unanyan2005}. 
It is worth mentioning that the mutual information $I_1$ is the $\textit{lower bound}$ 
to the quantum entanglement entropy $S$ \cite{Unanyan2005}. Note that  Shannon's mutual 
information $I_1$ (see Eq. (\ref{mutual2})) is equal to the R\'enyi entropy $S_n$ 
(see Eq. (\ref{Renyi from corr})) when $n=2$ \cite{Alcaraz2013}.  

According to Eqs. (\ref{HOLR}) and (\ref{HOLR2}), $K$ and $K^{\pm 1/2}$ matrices, for 
a translational invariant system, are Toeplitz matrices. Therefore, to compute the 
Shannon's classical mutual information Eq. (\ref{mutual1}), we need to compute the 
Teoplitz determinants. As shown by Fisher-Hartwig and proved later by Widom  \cite{Widom1973} (see Appendix. A) the 
asymptotic behavior of the Toeplitz determinants $\det(P_A)$ for the massless 
system i.e. Eq. (\ref{HOLR2}), with subsystem size $l\gg1$ is
\begin{eqnarray}\label{Widom}
\det P_A\propto l ^{\alpha^2/16}~.
\end{eqnarray}
It is also possible to apply the Fisher-Hartwig theorem to $X_A$ when $\alpha<2$. Then one can find the power law behavior 
\begin{eqnarray}\label{WidomX}
\det X_A\propto l ^{\alpha^2/16}~.
\end{eqnarray}
We have numerically calculated $X_A$ for $\alpha=2$, and found an agreement with the Eq. (\ref{WidomX}).
 
The Eqs. (\ref{Widom}) and (\ref{WidomX}), provide an explicit way to find 
the logarithmic behavior of the mutual information $I_1$  in terms of the 
system size. In the case where the system is very large and the subsystem has
 small size $l$, the mutual information $I_1$ can be obtained
\begin{eqnarray}\label{Scaling mutual unanyan small size}
I_1 = \frac{\alpha^2}{16}\ln l + c_0~.
\end{eqnarray} 
\begin{figure}[htp]
\begin{center}
\includegraphics[width=9cm,clip]{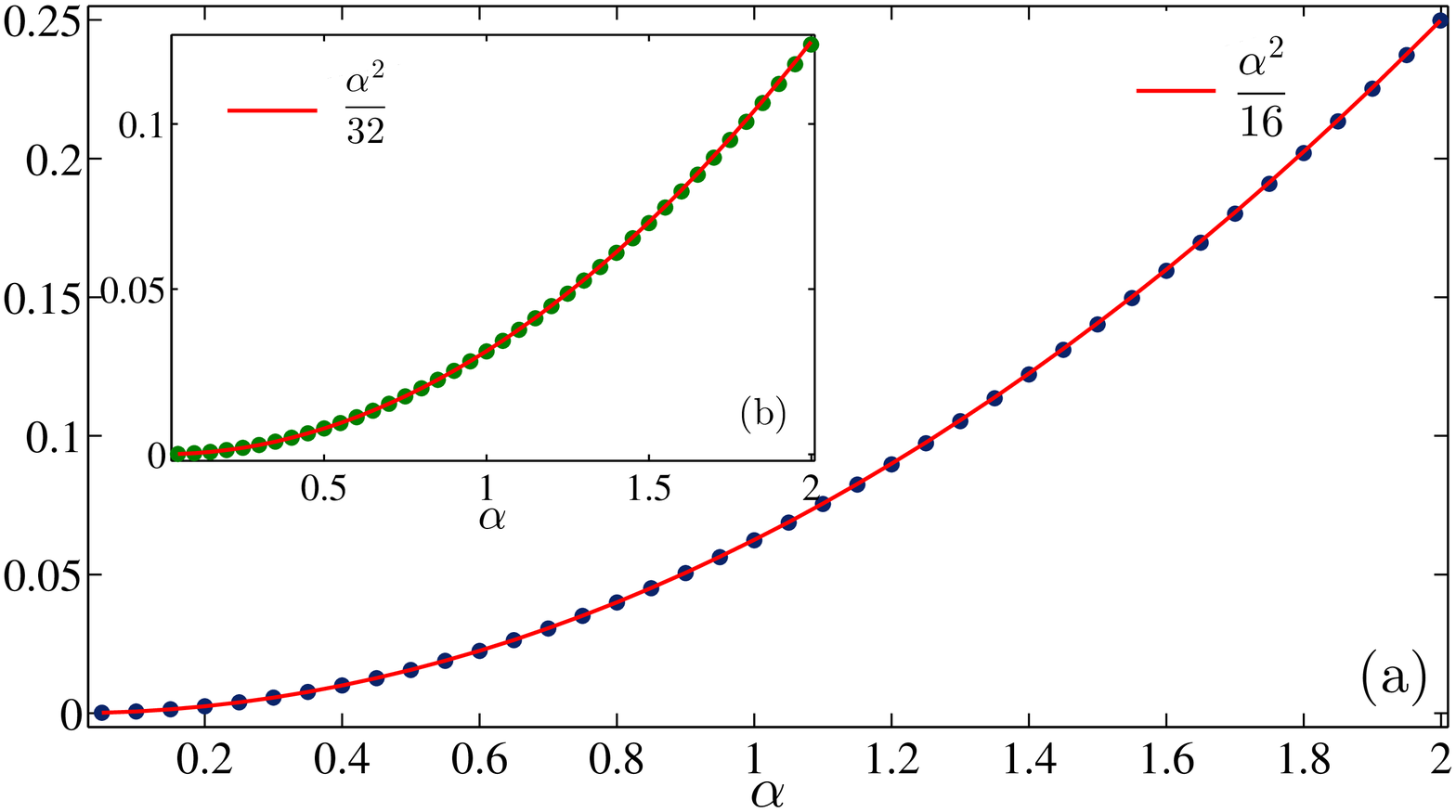}
\end{center}
\caption{(Color online) The prefactor of the logarithmic scaling of the mutual information $I_1$ for LRHO for a system with periodic boundary condition and configuration 
$\mathfrak{C_2}$. Inset: The same quantity for system with boundary and configuration $\mathfrak{C_4}$.}
\label{Figure:mutual pbc fbc wilczek}
\end{figure}
Numerical simulation results (see Fig. (\ref{Figure:mutual pbc small size 1})), in a wide range of $\alpha$, are in good agreement with the Eq. (\ref{Scaling mutual unanyan small size}), but 
when $1.5< \alpha <2 $ we observe small discrepancy in the numerical results. 
The reason of this discrepancy is unclear to us.

Here we also focus on the other definition considered by Cramer \textit{et al.} \cite{Cramer2006} 
to evaluate Shannon's mutual information. They determined the classical
Shannon entropy of the total lattice $S_{(A+B)}$, as well as the
entropy $S_A$ and $S_B$ determined by the reduced densities describing the 
two regions $A$ and $B$, respectively. The classical Shannon entropy for harmonic 
oscillator at finite temperature $T=1/\beta$ is 
\begin{eqnarray}
S_{\oplus} = -\frac{1}{2}\ln \det \left( K\vert _ {\oplus}\right)^{-1}+v(\oplus) \ln \frac{2\pi}{\beta} +v(\oplus)~,
\end{eqnarray}
where $\oplus\in\lbrace A,B,(A+B)\rbrace$ and $K\vert_\oplus$ denotes the $K$ matrix 
associated with the corresponding region $\oplus$ and $v$ is the size of the region. 
Then for a hamiltonian  of the form Eq. (\ref{harmonicOsc}) one can compute 
the Shannon's mutual information by the following formula:
\begin{eqnarray}\label{mutualeisert}
I_2 &= \frac{1}{2}\ln \frac{(\det K\vert _A)(\det K\vert _B)}{\det K}\nonumber \\ 
& = \frac{1}{2}\ln (\det K\vert _A K^{-1}\vert _A)~,
\end{eqnarray}
where $K^{-1}\vert_A$ denotes the $K^{-1}$ matrix associated with the interior region $A$. It is worth mentioning
 that the mutual information $I_2$ is independent of temperature.

Using Fisher-Hartwig theorem one can get the asymptotic behavior of the Toeplitz determinants 
$\det(K\vert_A)$ for the massless system i.e. Eq. (\ref{HOLR}) as
\begin{eqnarray}\label{WidomK}
\det K\vert_A \propto l ^{\alpha^2/4}~.
\end{eqnarray}

We will now discuss our numerical calculations. First suppose very large system and very small 
subsystem size (configuration $\mathfrak{C_1}$). In order to compute the mutual information 
we have numerically calculated the $K\vert _A$ and $K^{-1}\vert _A$ matrices. Then we calculated 
the eigenvalues of the matrix product $K\vert _AK^{-1}\vert _A$ and we measured the mutual information
 $I_2$ by the Eq. (\ref{mutualeisert}). Our results show the mutual information for LRHO increases 
logarithmically with the subsystem size as
\begin{eqnarray}\label{Scaling mutual eisert}
I_2 = \frac{\alpha^2}{8}\ln l + c_0~.
\end{eqnarray}
We then measured the prefactor of the logarithm and our results are shown in 
the Fig. (\ref{Figure:mutual pbc small size 2}). 
 
As we shall discuss in the next sections, it is easy
 to extend our numerical computation to 
general configurations of systems and sub-systems i.e. the configurations $\mathfrak{C_2}$, 
$\mathfrak{C_3}$ and $\mathfrak{C_4}$.   

\subsection{Finite systems}

Here we focus on the effect of the finite size system on the mutual information. Hence we shall 
first consider the mutual information $I_1$. Consider the case when the system has size $L$ and 
the subsystem has size $l=L/2$ (configuration $\mathfrak{C_2}$ and $\mathfrak{C_4}$). 
The mutual information $I_1$ for systems with size $L$ and subsystem size $l=L/2$ with periodic 
boundary condition (configuration $\mathfrak{C_2}$) follows \cite{Unanyan2005}:
\begin{eqnarray}\label{Scaling mutual unanyan PBC}
I_1 = \frac{\alpha^2}{16}\ln L + c_0,
\end{eqnarray}
where $\alpha$ is the scaling exponent for the LRHO and $L$ is the size of the 
system and $c_0$ is the non universal constant \cite{Unanyan2005}. 

Then consider the case with configuration $\mathfrak{C_4}$. In this case the mutual information $I_1$ follows
\begin{eqnarray}\label{Scaling mutual unanyan FBC}
I_1 = \frac{\alpha^2}{32}\ln L + c_0.
\end{eqnarray}
The numerical results of the prefactor of the logarithmic scaling Eqs. 
(\ref{Scaling mutual unanyan PBC}) and (\ref{Scaling mutual unanyan FBC}) for 
various $\alpha$'s are displayed in Fig. (\ref{Figure:mutual pbc fbc wilczek}). The agreement between the theoretical 
results given by Eqs. (\ref{Scaling mutual unanyan PBC}) and (\ref{Scaling mutual unanyan FBC}) 
and the numerical results is fairly good.

\begin{figure}[htp]
\begin{center}
\includegraphics[width=9cm,clip]{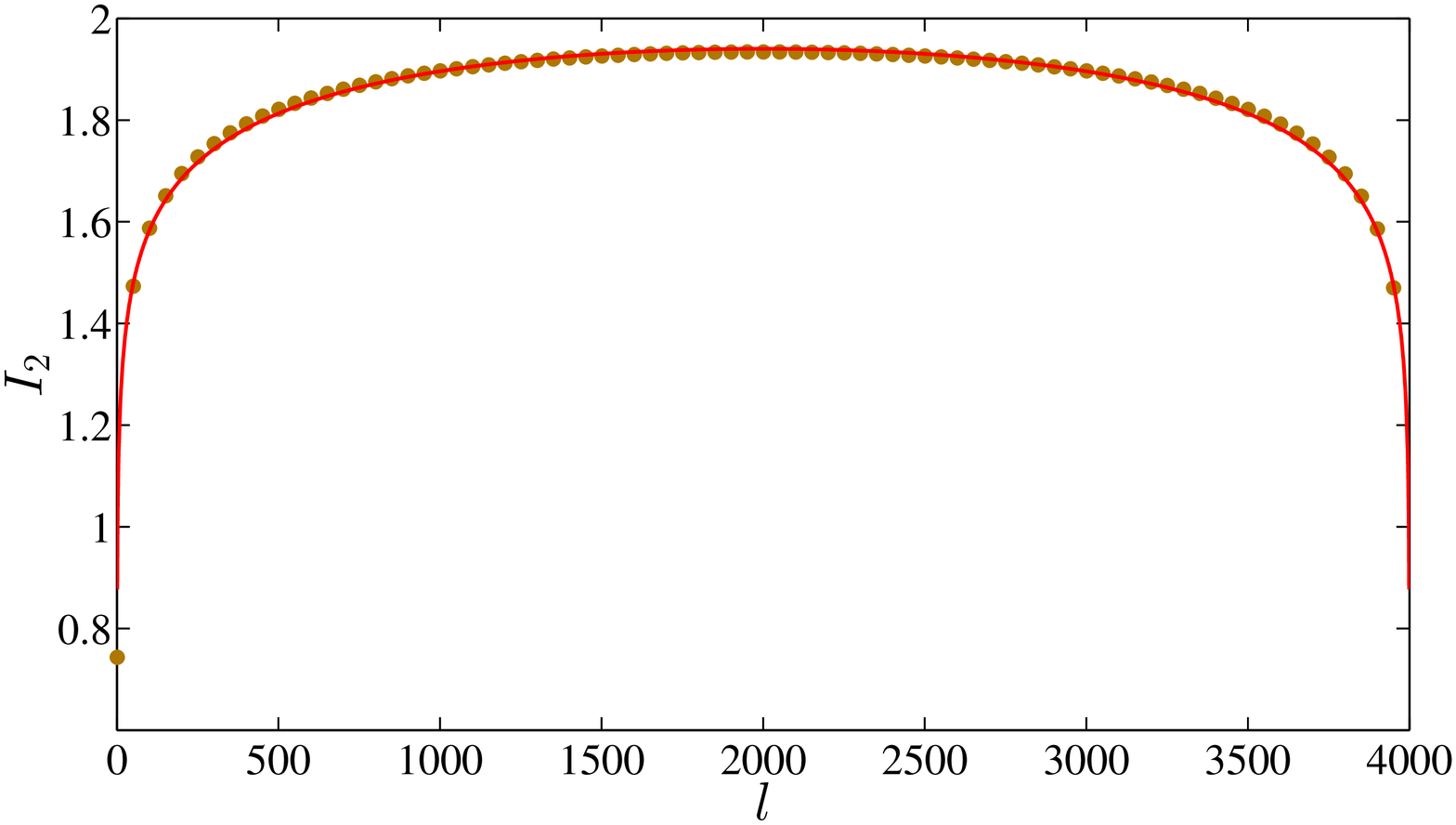}
\includegraphics[width=9cm,clip]{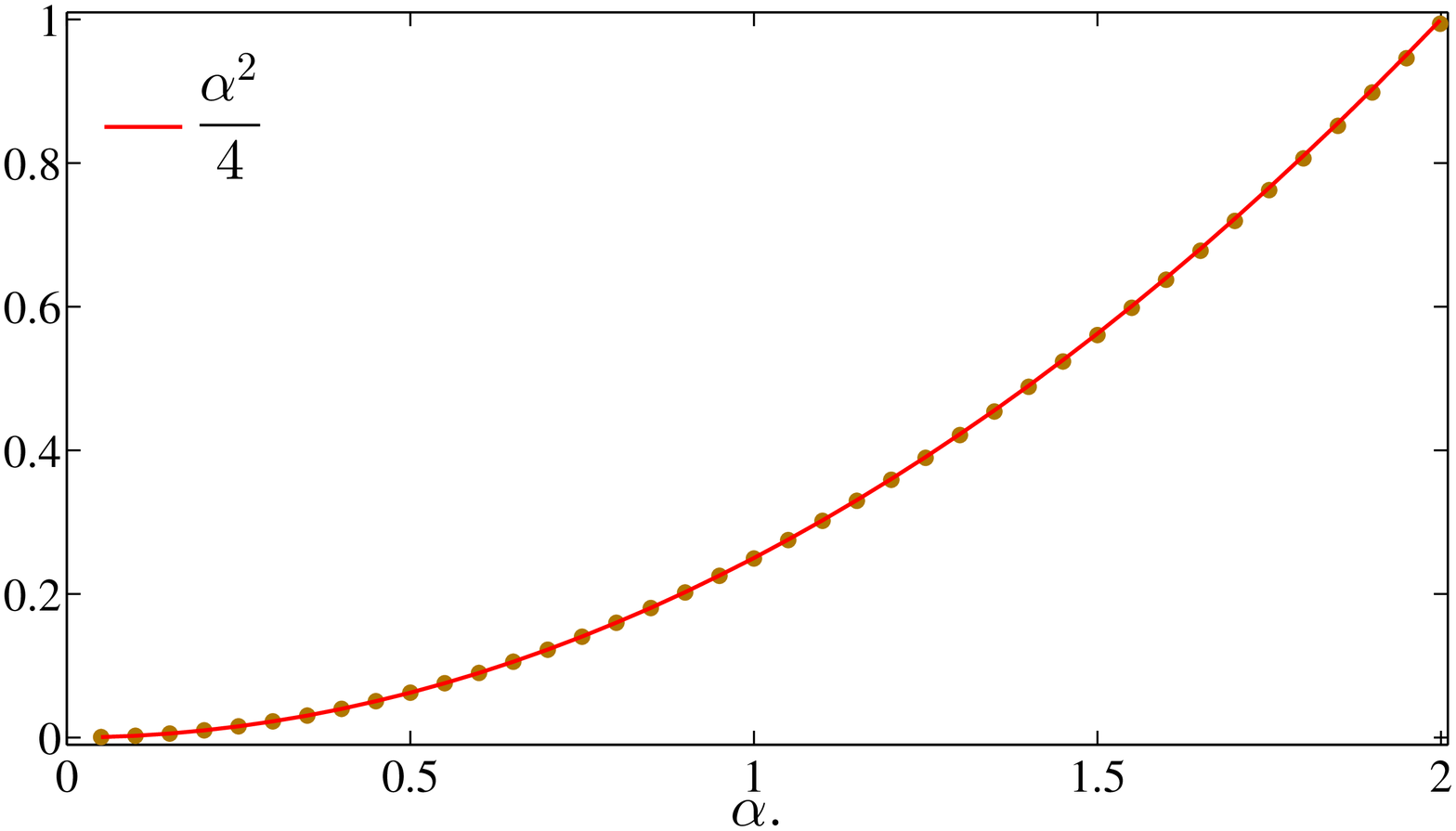}
\end{center}
\caption{(Color online) Top: Mutual information for LRHO ($\alpha=1.0$) with the 
configuration $\mathfrak{C_2}$. The solid line represents $I_2 =\frac{1}{8}\ln (l(L-l)) + c_0^\prime$. Bottom:
 The prefactor of the logarithmic scaling of the mutual information $I_2$ for LRHO for a system with periodic boundary condition and configuration $\mathfrak{C_2}$ when $l=L/2$.}
\label{Figure:mutual pbc fbc wilczek FSS}
\end{figure}

Now we are interested to find the mutual information $I_2$ for systems with finite size. 
First consider configuration $\mathfrak{C_2}$, when the subsystem has size $1<l<L/2$ and 
the system has periodic boundary condition. Finite size effects bend 
down the mutual information when the size of the sub-system approaches
half of the system size. 

Recall from Eq. (\ref{WidomK}) that $\det K\vert_A \propto l ^{\alpha^2/4}$ and $\det K\vert_B \propto (L-l) ^{\alpha^2/4}$ for a subsystem of length $l$ in a finite system of length $L$ with periodic boundary condition. It is then natural to expect that the mutual information $I_2$ (see Eq. (\ref{mutualeisert})) for systems with finite size, obeys the following formula: 
\begin{eqnarray}\label{mutual for finite eisert}
I_2 = \frac{\alpha^2}{8}\ln (l(L-l)) + c_0^\prime~.
\end{eqnarray}
We notice that the logarithmic scaling Eq. (\ref{Scaling mutual eisert}) can be 
recovered from Eq. (\ref{mutual for finite eisert}) for $l\ll L$. The numerical computation 
of the mutual information $I_2$ in this  case can be easily achieved using equation (\ref{mutualeisert}). 
The results are shown in Fig. (\ref{Figure:mutual pbc fbc wilczek FSS})
 obtaining excellent agreement between the numerical data and the Eq. (\ref{mutual for finite eisert}). 
\begin{figure}[htp]
\begin{center}
\includegraphics[width=9cm,clip]{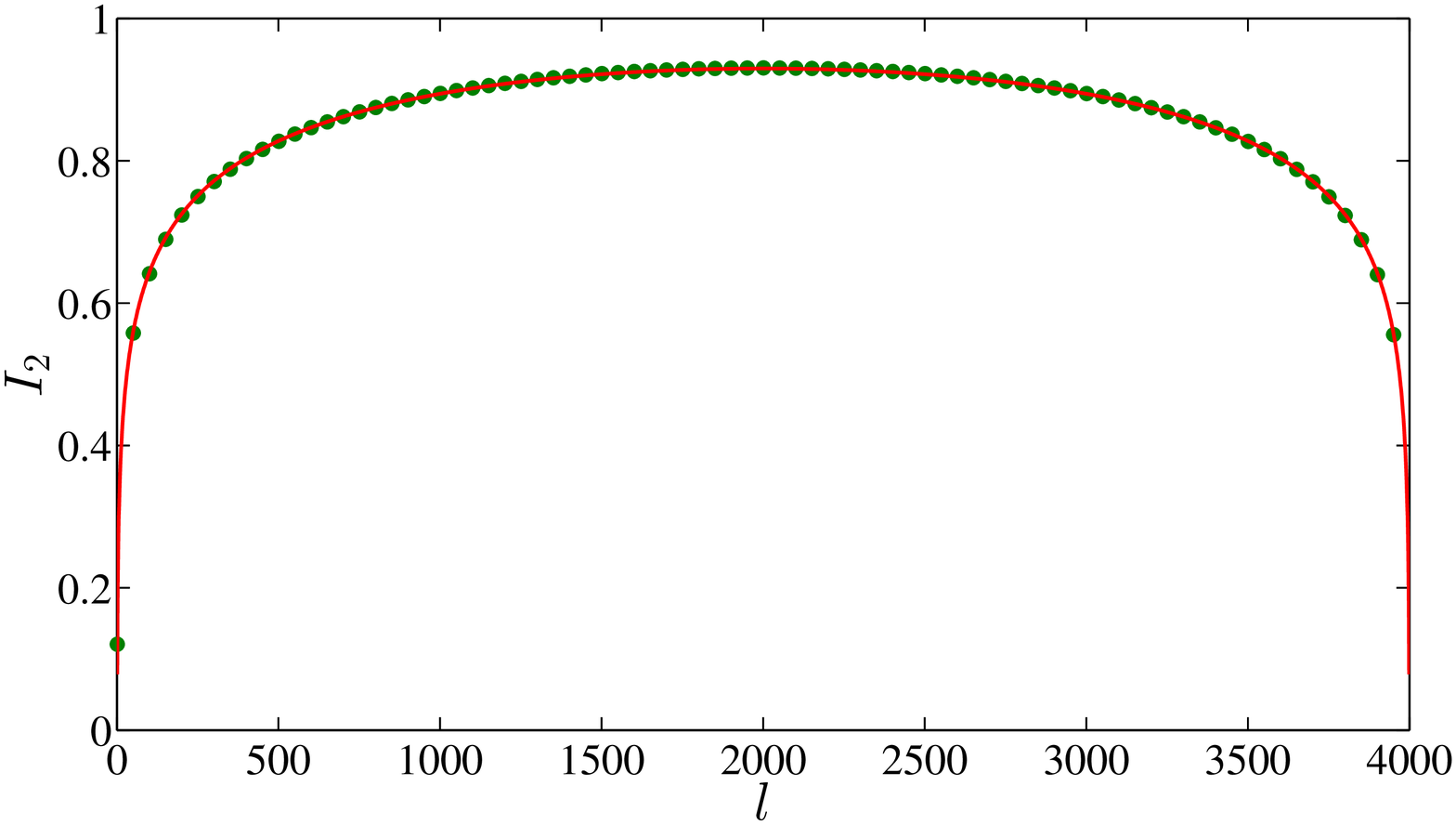}
\includegraphics[width=9cm,clip]{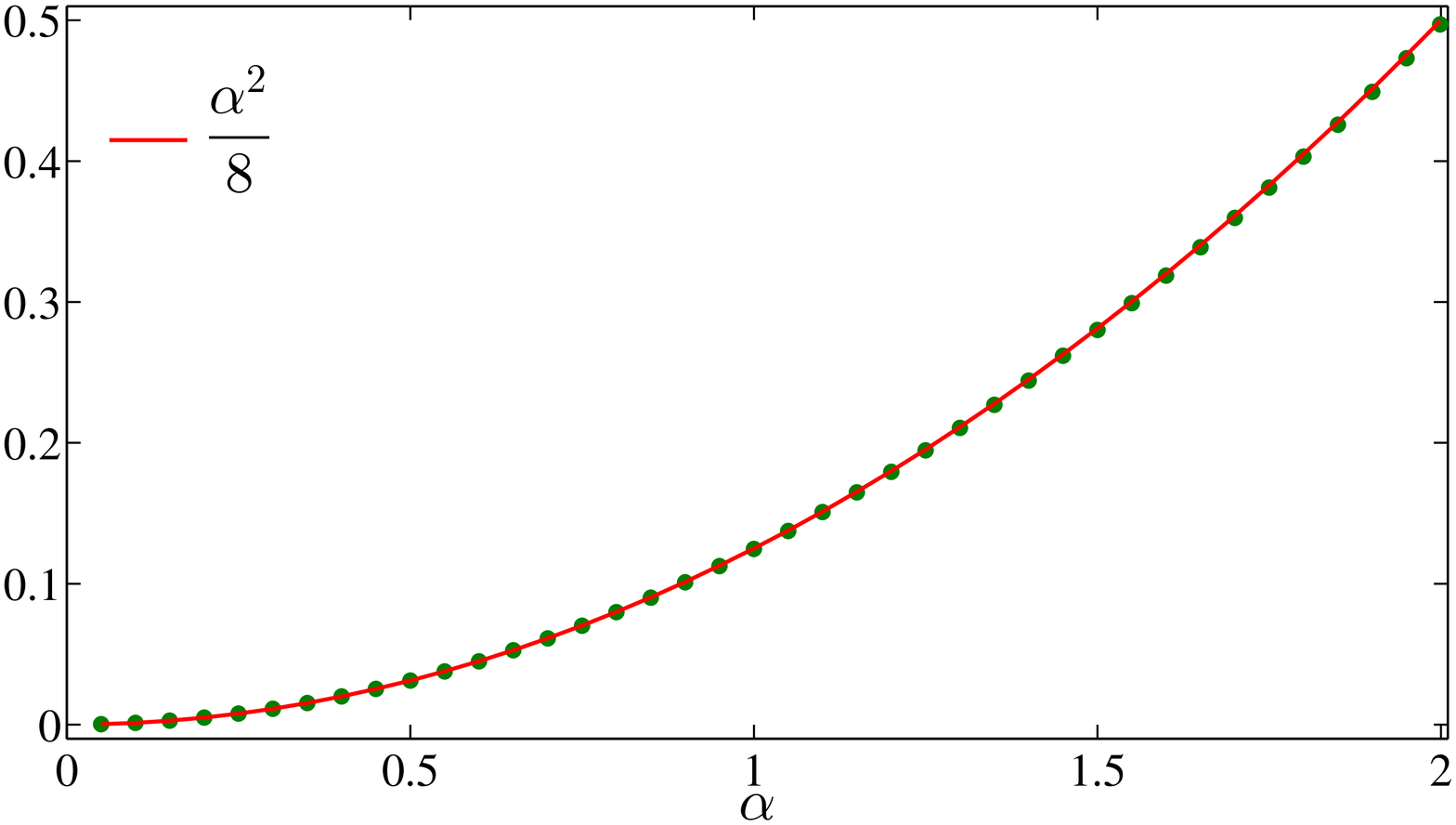}
\end{center}
\caption{(Color online) Top: Mutual information for LRHO ($\alpha=1.0$) with the configuration $\mathfrak{C_3}$.
 The solid line represents $I_2 =\frac{1}{8}\ln (l(L-l))-\frac{1}{8}\ln (L)+ c_0^\prime$. Bottom:
 The prefactor of the logarithmic scaling of the mutual information $I_2$ for LRHO  with configuration $\mathfrak{C_4}$.}
\label{Figure:mutual FSS}
\end{figure}
The mutual information $I_2$ for a system with periodic boundary condition 
and subsystem with size $l=L/2$ changes logarithmically as
\begin{eqnarray}\label{mutual for finite eisert L/2}
I_2 = \frac{\alpha^2}{4}\ln L + c_0^\prime~.
\end{eqnarray}
The numerical results are shown in Fig. (\ref{Figure:mutual pbc fbc wilczek FSS}). 
We obtain good agreement with the available theoretical prediction.

\begin{figure}[htp]
\begin{center}
\includegraphics[width=9cm,clip]{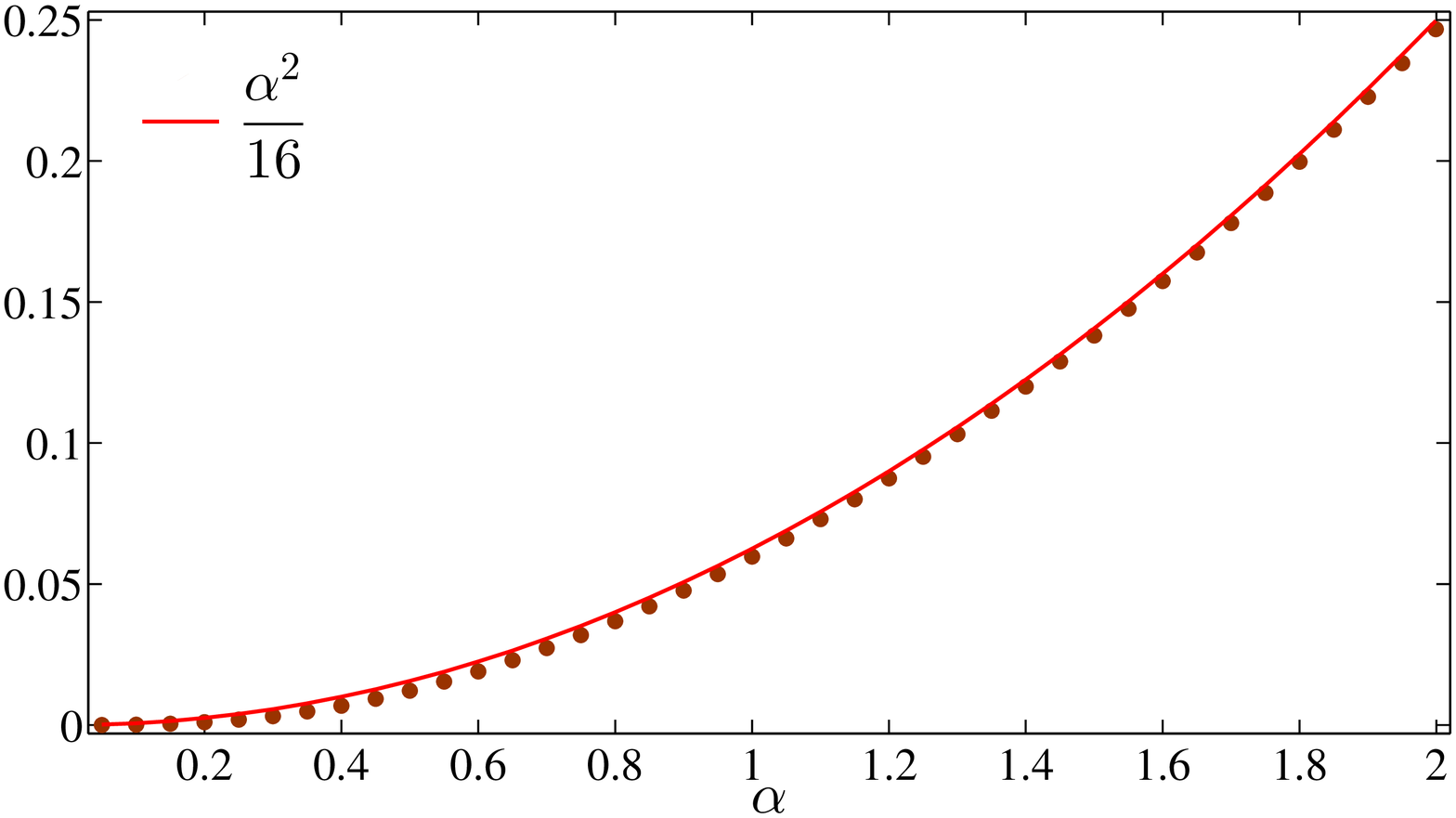}
\includegraphics[width=9cm,clip]{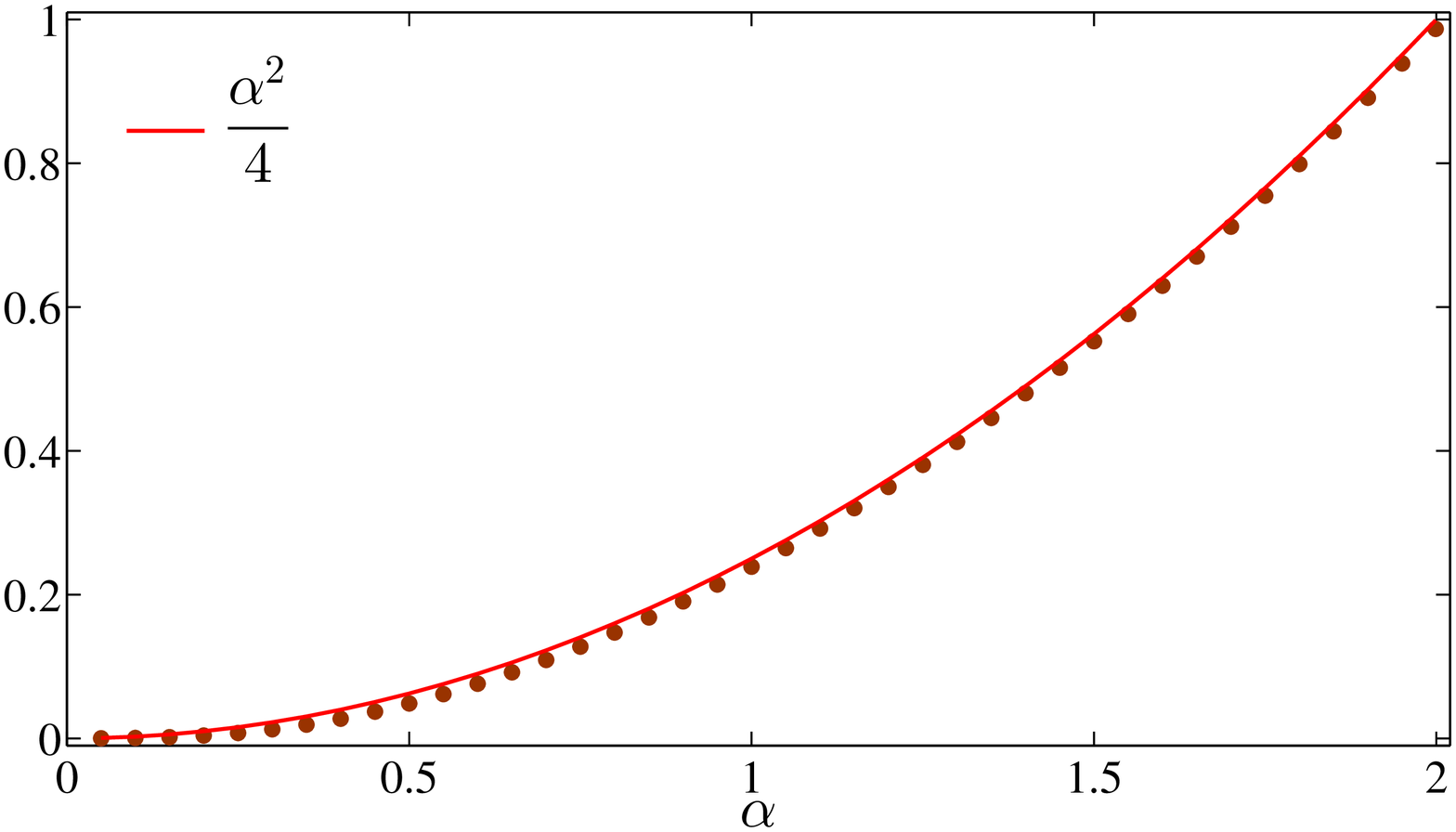}
\end{center}
\caption{(Color online) Top: The prefactor of the logarithmic scaling of the mutual information $I_1$ 
for massive LRHO with configuration $\mathfrak{C_5}$. Bottom: The same quantity for the mutual information $I_2$.}
\label{Figure:mutual pbc massive}
\end{figure}

We also studied  $I_2$ (see Eq. (\ref{mutualeisert})) for 
the long range harmonic oscillator with configuration $\mathfrak{C_3}$. Here, we 
examine the behavior of the mutual information $I_2$ for harmonic oscillator with 
long-range interaction, when the system has boundary. The boundary breaks translational 
symmetry and it is not therefore possible to use the method followed in \cite{Unanyan2005}. 
However, it is possible to find the $K$ matrix and then we can use  numerical 
diagonalization of the matrix $K\vert_A K^{-1}\vert_A$ to find the eigenvalues and 
calculate the mutual information $I_2$. In this case we observe that  
\begin{eqnarray}\label{mutual for finite with boundary eisert}
I_2 = \frac{\alpha^2}{8}\ln (l(L-l))-\frac{\alpha^2}{8}\ln(L) + c_0^\prime~,
\end{eqnarray}
where in our numerical simulations we found good agreement with our prediction 
(see Fig. (\ref{Figure:mutual FSS})). It is interesting to note that the mutual 
information $I_2$ for LRHO with configuration $\mathfrak{C_4}$ grows logarithmically with $L$;
\begin{eqnarray}\label{Scaling mutual boundary}
I_2 = \frac{\alpha^2}{8}\ln L + c_0~,
\end{eqnarray}
where this simple behavior is expected from the Eq. (\ref{mutual for finite with boundary eisert}) 
when $l=L/2$. In Fig. (\ref{Figure:mutual FSS}) we show our numerical results for the 
prefactor of the logarithmic term of the mutual information.
      
\subsection{Massive systems}

In this subsection we consider massive LRHO with $M>0$ 
(configuration $\mathfrak{C_5}$). For the massive case, we study the behavior of the 
mutual information $I_1$ and $I_2$ numerically. It is interesting to note that the mutual information $I_1$ and $I_2$ increase logarithmically
 with the mass  and obey the following formula:
\begin{eqnarray}\label{Scaling mutual massive}
I_1 = -\frac{\alpha^2}{16}\ln M, \hspace{1cm}I_2 = -\frac{\alpha^2}{4}\ln M~.
\end{eqnarray}   
In Fig. (\ref{Figure:mutual pbc massive}) we report 
the results of the simulation of the mutual information $I_1$ and $I_2$ for 
the massive LRHO, where we calculated the prefactor of the logarithm  which is in  good agreement with the 
 Eq. (\ref{Scaling mutual massive}).

\subsection{Generalized singular Toeplitz matrices}

In this subsection we generalize our results to the general Toeplitz matrices that we have studied in the
 section \ref{Generalized entropy}. For $I_1$ the discussion follows the argument given in \cite{Unanyan2005}
which is based on the Fisher-Hartwig theorem. It is very simple to see that since $P_A$ and $X_A$ are Topelitz matrices for the 
$\alpha_r<2$ one can simply get the following results for the prefactor of the logarithm of the 
mutual information $\tilde{c_{I_1}}$ of the subsystem

\begin{eqnarray}\label{mutual Toeplitz I1}
\begin{split}
&\tilde{c}_{I_1}(\alpha_1,\alpha_1,\alpha_3,\alpha_3,\dots,\alpha_R)=\\ 
&\tilde{c}_{I_1}(\alpha_1,\alpha_1,\dots,0)+\tilde{c}_{I_1}(0,0,\alpha_2,\alpha_2,\dots,0)+\dots\\
&+\tilde{c}_{I_1}(0,0,\dots,\alpha_{R-1},\alpha_{R-1}).
\end{split}
\end{eqnarray}

Similar result has been already announced in \cite{Unanyan2005} for the mutual information of a periodic system with half of
the system as the subsystem for $\alpha=even$. We numerically checked the above result in table ~\ref{tab2}. It is worth mentioning
that the prefactors are independent of $b(q)$ and $q_r$'s.

\begin{table}[htp]
\begin{center}
\begin{tabular}{l*{15}{c}r}    
\hline\hline
$q_1$ & $q_2$ & $q_3$ & $q_4$ & $q_5$& $q_6$& $\alpha_1$ & $\alpha_2$ & $\alpha_3$ & $\alpha_4$ & $\alpha_5$& $\alpha_6$&  ${\tilde{c}_{I_1}}/{2}$\\ \hline
$\frac{\pi}{3}$& $\frac{-\pi}{3}$ & 0 & 0 & 0  & 0 & 1 & 1 & 0  & 0  & 0  & 0  &  0.059(0.001) \\ 
\hline
$\frac{\pi}{6}$& $\frac{-\pi}{6}$ & 0 & 0 & 0  & 0 & 1 & 1 & 0  & 0  & 0  & 0  &  0.059(0.001) \\ 
\hline
0 & 0 & $\frac{\pi}{3}$ & $\frac{-\pi}{3}$ & 0 & 0  & 0 & 0 & 1.5 & 1.5  & 0 & 0  &  0.13(0.01) \\  
\hline     
0& 0  & 0 & 0 & $\frac{\pi}{3}$  & $\frac{-\pi}{3}$ & 0 & 0 & 0  & 0  & 2  & 2  &  0.24(0.01) \\  
\hline  
$\frac{\pi}{3}$ & $\frac{-\pi}{3}$ & $\frac{\pi}{6}$ & $\frac{-\pi}{6}$ & 0  & 0 & 1 & 1 & 1.5  & 1.5  & 0  & 0  &  0.19(0.01) \\ 
\hline
$\frac{\pi}{3}$ & $\frac{-\pi}{3}$ & $\frac{\pi}{6}$ & $\frac{-\pi}{6}$ & 0  & 0 & 1 & 1 & 1.5  & 1.5  & 2  & 2  &  0.44(0.02) \\  
\hline     
$\frac{\pi}{3}$ & $\frac{-\pi}{3}$ & $\frac{\pi}{6}$ & $\frac{-\pi}{6}$ & $\frac{\pi}{4}$  & $\frac{-\pi}{4}$ & 1 & 1 & 1.5  & 1.5  & 2  & 2  &  0.44(0.02) \\ 
\hline\hline
    \end{tabular}
\end{center}
\caption{\label{tab2} Numerical values of the prefactor ${\tilde{c}_{I_1}}$ for different values of
$\alpha_r$ and $q_r$.}
\end{table}

It is not difficult to show that the same formula is also valid for the cases with boundary.

Finally we should mention that the above argument works perfectly also for $I_2$. The results of
 the numerical calculations are shown in the table ~\ref{tab3} for a the large system with the small subsystem $A$.
\begin{table}[htp]
  \begin{center}
{\begin{tabular}{l*{15}{c}r}    
\hline\hline
$q_1$ & $q_2$ & $q_3$ & $q_4$ & $q_5$& $q_6$& $\alpha_1$ & $\alpha_2$ & $\alpha_3$ & $\alpha_4$ & $\alpha_5$& $\alpha_6$&  $\tilde{c}_{I_2}/2$\\ \hline
$\frac{\pi}{3}$& $\frac{-\pi}{3}$ & 0 & 0 & 0  & 0 & 1 & 1 & 0  & 0  & 0  & 0  &  0.125(0.001) \\ 
\hline
$\frac{\pi}{6}$& $\frac{-\pi}{6}$ & 0 & 0 & 0  & 0 & 1 & 1 & 0  & 0  & 0  & 0  &  0.125(0.001) \\ 
\hline
0 & 0 & $\frac{\pi}{3}$ & $\frac{-\pi}{3}$ & 0 & 0  & 0 & 0 & 1.5 & 1.5  & 0 & 0  &  0.27(0.01) \\  
\hline     
0& 0  & 0 & 0 & $\frac{\pi}{3}$  & $\frac{-\pi}{3}$ & 0 & 0 & 0  & 0  & 2  & 2  &  0.48(0.01) \\  
\hline  
$\frac{\pi}{3}$ & $\frac{-\pi}{3}$ & $\frac{\pi}{6}$ & $\frac{-\pi}{6}$ & 0  & 0 & 1 & 1 & 1.5  & 1.5  & 0  & 0  &  0.39(0.01) \\ 
\hline
$\frac{\pi}{3}$ & $\frac{-\pi}{3}$ & $\frac{\pi}{6}$ & $\frac{-\pi}{6}$ & 0  & 0 & 1 & 1 & 1.5  & 1.5  & 2  & 2  &  0.88(0.02) \\  
\hline     
$\frac{\pi}{3}$ & $\frac{-\pi}{3}$ & $\frac{\pi}{6}$ & $\frac{-\pi}{6}$ & $\frac{\pi}{4}$  & $\frac{-\pi}{4}$ & 1 & 1 & 1.5  & 1.5  & 2  & 2  &  0.88(0.02) \\ 
\hline\hline
    \end{tabular} }
\end{center}
\caption{\label{tab3} Numerical values of the prefactor ${\tilde{c}_{I_2}}$ for different values of
$\alpha_r$ and $q_r$.}
\end{table}

\begin{figure} [htp]
\centering
\includegraphics[width=9cm,clip]{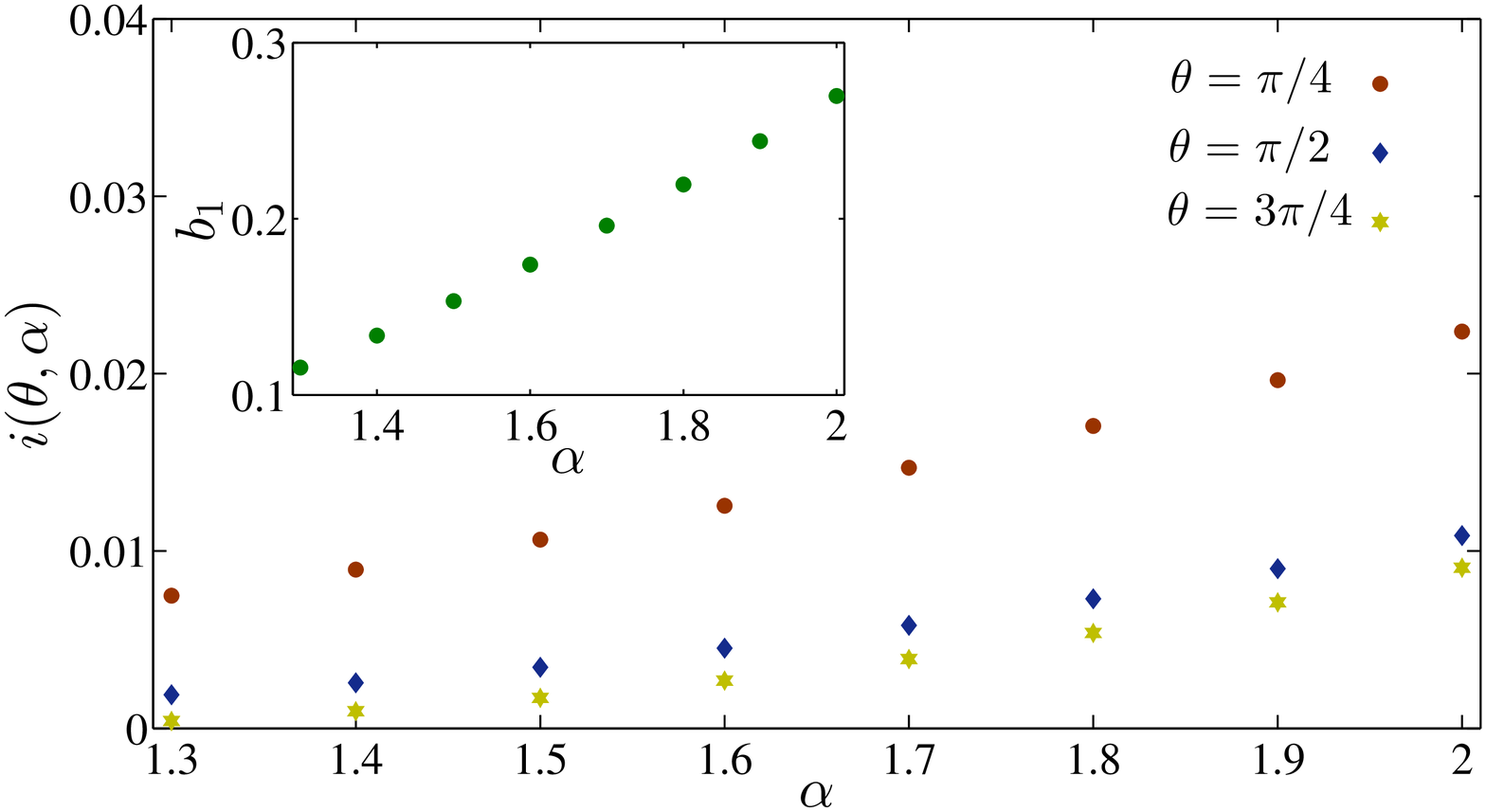}
\caption{(Color online) $i(\theta,\alpha)$ as a function of $\alpha$ for subsystems with different vertex angles $\theta$. \textit{Inset}:
the nonuniversal coefficient of the area term $b_1$ with respect to different values of $\alpha$. }
\label{Figure:Mutual_2d_area_and_log_coeff}
\end{figure}

\subsection{Two dimensions: area law and logarithmic term for polygonal region}

In this section we study the behavior of the Shannon mutual information in two dimensions. Since $I_1$ is nothing exept the $n=2$ R\'enyi entanglement entropy
one  just expect that the equation (\ref{EE polygon}) be valid also for this case. In this section we mostly concentrate on $I_2$. We first confirmed
that the area law is valid also in this case for different values of $\alpha=2,1.5$ and $1$. Then we checked the effect of sharp corner as we did in the case
of entanglement entropy. The best fit for the data is

\begin{eqnarray}\label{mutual polygon}
I_{2}(\theta)=b_0+b_1 L+b_{-1}L^{-1}+b_{-2}L^{-2}-i(\theta,\alpha)\log L,
\end{eqnarray}
where $\theta$ as before is the vertex angle. The coefficient of the area term is a nonuniversal quantity and  increases with the $\alpha$, 
see Fig. (\ref{Figure:Mutual_2d_area_and_log_coeff}).
similar to what we had in the case of the entanglement entropy the coefficient of the logarithm is a universal function and increases with
$\alpha$ and decreases with $\theta$, see Fig. (\ref{Figure:Mutual_2d_area_and_log_coeff}). It is worth mentioning that we also calculated the same quantity
for $I_1$ and we found that $i(\theta,\alpha)$ is $\frac{1}{4}$ of the result for $I_2$.

\section{Conclusions and Discussions}

In this paper we studied quantum entanglement entropy of coupled long-range harmonic oscillators. We showed that the von Neumann
and R\'enyi entanglement entropy of a subsystem of an infinite system changes logarithmically with the subsystem size which the prefactor
is dependent to the fractional power of the interaction $\alpha$. We also studied the same quantities in the presence of different kinds 
of boundary conditions and found that the entanglement entropy  changes logarithmically with the subsystem size but with a prefactor which is different 
from the case without a boundary. The prefactor is interestingly the same as the prefactor coming from the massive case. Having the above results we concluded that 
there are just two universal prefactors in our system. Later we extended our results to the finite temperature case and found $T^{\frac{2}{\alpha}}$ dependence 
of the entanglement entropy to the temperature. Our main result was studying the universality of our results
by changing the interactions. For example we considered long-range HO plus short-rang HO and found that the short-range interaction does not have any effect as far as $\alpha<2$. For $\alpha>2$ the result is the same as the short-range interaction. We also showed that one can change some other
parameters in the interaction and get always the same results. We generalized our findings by studying general singular Toeplitz like couplings
which in this case we showed that one can calculate the entanglement entropy by just having the results for the simple cases that we have studied.
Although in this case we were able to prove the result for the $n=2$ R\'enyi case, proving it for the von Neumann entanglement entropy is far from obvious. We also generalized 
our findings to two dimensional cases and showed that despite the long-range nature of the couplings the area law is valid in this case. In addition
we showed that universal logarithmic terms will appear if we consider regions with sharp corner in our system. 
Finally we also studied mutual shannon entropy in our system. We used two definitions; one coming from purely classical considerations and the other comes
from using the ground state of the quantum system as the source of probabilities. We showed that the latter case is actually equal to the $n=2$
R\'enyi entanglement entropy and one can calculate many things analytically by using Fisher-Hartwig theorem for Toeplitz matrices. We also provided
many simple exact results by using the same method. The generalization to the singular Toeplitz matrices is immediate in these two cases and one can
prove that the decomposition mentioned in the case of von Neumann entropy is valid also in this case. There are many other directions that one can extend our work
among the immediate ones one can call the study of our system in the presence of the quantum quench, the other direction is studying the entanglement entropy
of excited states. Another important study can be investigating the entanglement entropy of 
long-range Ising model in the mean field regim where one can relate it to the field theory that we have studied in this paper.
We hope to be able to come back to some of these questions in future.
 \section*{Acknowledgments}
MGN thanks R. Metzler for supports and A. Chechkin for helpful discussions. MGN acknowledges financial support from University of Potsdam. MAR thanks FAPESP for financial support.

\appendix
\section{Appendix: Fisher-Hartwig theorem}

The Fisher-Hartwig conjecture which is proved later by Widom \cite{Widom1973} is about the asymptotic behavior of the determinants of a certain class of
Toeplitz matrices. The singular Toeplitz matrices have the following form
\begin{eqnarray}\label{Toeplitz Matrics}
K_{l,m}= - \int _{0}^{2\pi} \frac{dq}{2\pi}e^{iq(l-m)} {b(q)\phi(q-q_r)},
\end{eqnarray}
where $b(q):S^1\to \mathcal{C}$ is a smooth non-vanishing function with zero winding number and 
\begin{eqnarray}\label{u function Toeplitz}
\phi(q)&=&\prod_{r=1}^{R}u(\alpha_r,q)t(\beta_r,q),\\
u(\alpha,q)&=&(2-2\cos q)^{\frac{\alpha}{2}},\hspace{1cm}\text{Re}\alpha>-1\\
t(\beta,q)&=&\exp[-i\beta(\pi-q)],\hspace{1cm} 0<q<2\pi.
\end{eqnarray}
Fisher and Hartwig conjectured that the determinant of the matrix $K$ follows
\begin{eqnarray}\label{fisher Hartwig conjecture}
D_n[K]\sim E G^n(b) n^{\sum_r (\frac{\alpha^2_r}{4}-\beta_r^2)} ,
\end{eqnarray}
where $E$ is a constant and $G(b)=\exp(\frac{1}{2\pi}\int_0^{2\pi}\log b(q)dq)$. In our study we took the cases with $\beta_r=0$
, however, we believe that generalizations to $\beta\neq 0$ are straightforward.

\end{document}